\documentclass[aps,prd,twocolumn,showpacs,amsmath,amssymb]{revtex4}%
\usepackage{graphicx}% Include figure files
\usepackage{epsfig}
\usepackage{graphics,color}
\usepackage{amssymb}
\usepackage{amsbsy}
\usepackage{bm}

\newcommand{\vev}[1]{\langle#1\rangle}
\newcommand{\vect}{\left ( \begin{array}{c}}
\newcommand{\evect}{\end{array} \right )}

\newcommand{\beq}{\begin{equation}}
\newcommand{\eeq}{\end{equation}}
\newcommand{\bea}{\begin{eqnarray}}
\newcommand{\eea}{\end{eqnarray}}
\newcommand{\ba}{\begin{array}}
\newcommand{\ea}{\end{array}}

\newcommand{\bef}{\begin{figure}}
\newcommand{\eef}{\end{figure}}
\newcommand{\bce}{\begin{center}}
\newcommand{\ece}{\end{center}}

\def\fsl#1{\setbox0=\hbox{$#1$}                 % set a box for #1
   \dimen0=\wd0                                 % and get its size
   \setbox1=\hbox{/} \dimen1=\wd1               % get size of /
   \ifdim\dimen0>\dimen1                        % #1 is bigger
      \rlap{\hbox to \dimen0{\hfil/\hfil}}      % so center / in box
      #1                                        % and print #1
   \else                                        % / is bigger
      \rlap{\hbox to \dimen1{\hfil$#1$\hfil}}   % so center #1
      /                                         % and print /
   \fi}
\newcommand{\gsim}{\mathrel{\hbox{\rlap{\lower.55ex \hbox {$\sim$}}
                   \kern-.3em \raise.4ex \hbox{$>$}}}}
\newcommand{\lsim}{\mathrel{\hbox{\rlap{\lower.55ex \hbox {$\sim$}}
                   \kern-.3em \raise.4ex \hbox{$<$}}}}

\begin{document}

\title{QCD phase diagram at high temperature and density}
\author{Mei Huang$^{1,2}$ \footnote{huangm@ihep.ac.cn}}
\affiliation{$^{1}$Institute of High Energy Physics, Chinese Academy
of Sciences, Beijing, China \\
$^{2}$ Theoretical Physics Center for Science Facilities, Chinese
Academy of Sciences, Beijing, China}

\begin{abstract}
This article reviews recent progress of QCD phase structure,
including color superconductor at high baryon density and strongly
interacting quark-gluon plasma (sQGP) at high temperature created
through relativistic heavy ion collision. A brief overview is given
on the discovery of sQGP at RHIC. The possibility of locating the
critical end point (CEP) from the property of bulk viscosity over
entropy density is discussed. For the phase structure at high baryon
density, the status of the unconventional color superconducting
phase with mismatched pairing is reviewed. The chromomagnetic
instability, Sarma instability and Higgs instability in the gapless
color superconducting phase are clarified.
\end{abstract}

\maketitle

\section{Introduction}
\label{secIntro}

Quantum Chromodynamics (QCD) is an asymptotically free theory
\cite{asymptotic} and regarded as the fundamental theory of quarks
and gluons. At very high energies, interaction forces become weak,
thus perturbation calculations can be used. The perturbative QCD
predictions have been extensively confirmed by experiments, while
QCD in the non-perturbative regime is still a challenge to
theorists. The fundamental quarks and gluons of QCD have not been
seen as free particles, but are always confined within hadrons. It
is still difficult to construct the hadrons in terms of nearly
massless quarks and gluons. The observed baryon spectrum indicates
that the (approximate) chiral symmetry is spontaneously broken in
the vacuum. As a result, the eight pseudoscalar mesons $\pi$, $K$
and $\eta$ are light pseudo-Nambu-Goldstone bosons, and the
constituent quark obtains dynamical mass, which contributes to the
baryon mass. Besides conventional mesons and baryons, QCD itself
does not exclude the existence of the non-conventional states such
as glueballs, hybrid mesons and multi-quark states
\cite{Klempt:2004yz}.

\begin{figure}
\includegraphics[width=7cm]{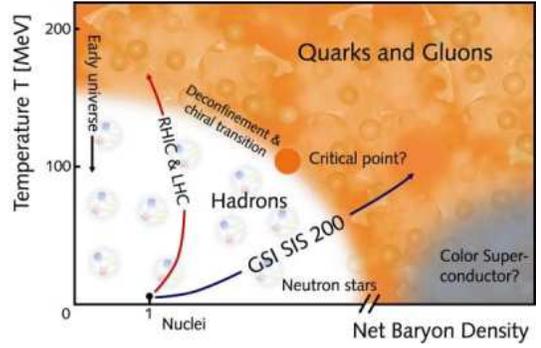}
\caption{ QCD phase diagram at finite temperature and baryon
density.} \label{fig-phaseDiagram}
\end{figure}

Since 1970s, people have been interested in QCD at extreme
conditions. It is expected that the chiral symmetry can be restored,
and quarks and gluons will become deconfined at high temperatures
and/or densities \cite{Lee:1974ma,Collins:1974ky,Baym:1976yu,qgpT}.
Fig. \ref{fig-phaseDiagram} is the typical QCD phase diagram, which
shows the system is in deconfined quark-gluon plasma phase at high
temperature, and in color superconducting phase at high baryon
density.

Results from lattice show that the quark-gluon plasma (QGP) does
exist. For the system with zero net baryon density, the
deconfinement and chiral symmetry restoration phase transitions
happen at the same critical temperature \cite{Karsch}. At
asymptotically high temperatures, e.g., during the first
microseconds of the ``Big Bang", the many-body system composed of
quarks and gluons can be regarded as an ideal Fermi and Boson gas.
It is believed that the ``little Bang" can be produced at RHIC and
LHC. Recently, it was shown that the new state of matter produced at
RHIC is far away from the asymptotically hot QGP, but in a strongly
coupled regime. This state is called strongly coupled quark-gluon
plasma (sQGP)\cite{Shuryak-SQGP}. For most recent reviews about QGP,
e.g., see
Ref.~\cite{Rischke-r,Gyulassy-Mclerran,Jacobs-Wang,Heinz-review}.

Studying QCD at finite baryon density is the traditional subject of nuclear
physics. The behaviour of QCD at finite baryon density and low temperature is
central for astrophysics to understand the structure of compact stars, and
conditions near the core of collapsing stars (supernovae, hypernovae).
Cold nuclear matter, such as in the interior of a Pb nucleus, is
at $T=0$ and $\mu_B\simeq m_N = 940 {\rm MeV}$. Emerging from this point,
there is a first-order nuclear liquid-gas phase transition,
which terminates in a critical endpoint at a temperature
$\sim 10 {\rm MeV}$ \cite{liquid-gas-Tc}. If one squeezes matter further
and further, nuleons will overlap. Quarks and gluons in one nucleon can feel
quarks and gluons in other nucleons. Eventually, deconfinement phase
transition will happen. Unfortunately, at the moment, lattice QCD is facing
the ``sign problem" at nonzero net baryon densities. Our understanding at
finite baryon densities has to rely on effective QCD models. Phenomenological
models indicated that, at nonzero baryon density, the QGP phase and the hadron
gas are separated by a critical line of roughly a constant energy density
$\epsilon_{cr} \simeq 1 {\rm GeV}/{\rm fm}^3$ \cite{Heinz-Stocker}.

In the case of asymptotically high baryon density, the system is a
color superconducor. This was proposed by Frautschi \cite{Frautschi}
and Barrois \cite{Barrois}. Based on the Bardeen, Cooper, and
Schrieffer (BCS) theory \cite{BCS}, because there is a weak
attractive interaction in the color antitriplet channel, the system
is unstable with respect to the formation of particle-particle
Cooper-pair condensate in the momentum space. Detailed numerical
calculations of color superconducting gaps were firstly carried out
by Bailin and Love \cite{Bailin-Love}. They concluded that the
one-gluon exchange induces gaps on the order of 1 ${\rm MeV}$ at
several times of nuclear matter density. This small gap has little
effect on cold dense quark matter, thus the investigation of cold
quark matter lay dormant for several decades. It was only revived
recently when it was found that the color superconducting gap can be
of the order of $100~{\rm MeV}$ \cite{CS-1997}, which is two orders
larger than early perturbative estimates in Ref.~\cite{Bailin-Love}.
For this reason, the topic of color superconductivity stirred a lot
of interest in recent years. For review articles on the color
superconductivity, see for example, Refs.~\cite{CSC-review}.

In this article, I will focus on recent progress of sQGP created at
RHIC and color superconducting phase structure at intermediate
baryon density regime. The outline of this article is as follows: I
will give a brief overview on the discovery of sQGP at RHIC in Sec.
\ref{sec-sQGP}. Then introduce the status of the color
superconducting phase especially the gapless color superconducting
phase in Sec. \ref{sec-CSC}. At last, I will give a brief outlook in
Sec.~\ref{outlook}.

%%%%%%%%%%%%%%%%%%%%%%%%%%%%%%%%%%%%

\section{Strongly interacting quark-gluon plasma(sQGP)}
\label{sec-sQGP}

\subsection{Discovery of sQGP at RHIC}

Studying Quantum chromodynamics (QCD) phase transition and
properties of hot quark matter at high temperature has been the main
target of heavy ion collision experiments at the Relativistic Heavy
Ion collider (RHIC), the forthcoming Large Hadron Collider (LHC) and
FAIR at GSI.

The deconfined quark-gluon plasma, if it can be created through
heavy-ion collisions, is an intermediate state and cannot be
measured directly. In experiment, the detector can only measure the
freeze-out hadrons. In order to extract the property of the
intermediate state, hydrodynamics is often used to simulate the
evolution of the fluid.

The hydrodynamical equations of motion are the local conservation
laws of energy-momentum and net charge
\begin{equation}
\partial_{\mu}T^{\mu\nu}=0, ~~ \partial_{\mu}N_c^{\mu}=0 .
\end{equation}
In ideal hydrodynamics, the energy-momentum tensor takes the form of
\begin{equation}
T_{ideal}^{\mu\nu}=(\epsilon+p)u^{\mu}u^{\nu}-p g^{\mu\nu},
\end{equation}
with $u^{\mu}$ the flow velocity, $\epsilon, p$ the energy density
and pressure density, respectively.

In the Navier-Stokes hydrodynamics, the energy momentum tensor
decomposes into ideal and dissipative parts as
\begin{equation}
T_{NS}^{\mu\nu}=T_{ideal}^{\mu\nu}+\tau^{\mu\nu},
\end{equation}
with
\begin{equation}
\tau^{\mu\nu}=\eta(\nabla^{\mu}u^{\nu}+\nabla^{\nu}u^{\mu}-
\frac{2}{3}\triangle^{\mu\nu}\nabla_{\alpha}u^{\alpha}
 + \zeta \triangle^{\mu\nu}\nabla_{\alpha}u^{\alpha}.
\end{equation}
Where $\triangle^{\mu\nu}=g^{\mu\nu}-u^{\mu}u^{\nu},
\nabla^{\mu}=\triangle^{\mu\nu}\partial_{\nu}$, $\eta, \zeta$ are
the shear viscosity and bulk viscosity, respectively.

It was expected that deconfined quark matter formed at high
temperature should behave like a gas of weakly interacting
quark-gluon plasma (wQGP). The perturbative QCD calculation gives a
large shear viscosity in the wQGP with $\eta/s\simeq 0.8$ for
$\alpha_s=0.3$ \cite{Arnold-shear}. Therefore, it turned out as a
surprise that the RHIC data of elliptic flow $v_2$ can be described
very well by requiring a very small shear viscosity over entropy
density ratio $\eta/s$ \cite{Hydro, Hydro-Teaney}. Lattice QCD
calculation confirmed that $\eta/s$ for the purely gluonic plasma is
rather small and in the range of $0.1-0.2$ \cite{LAT-etas}.

It is now believed that the system created at RHIC is a strongly
coupled quark-gluon plasma (sQGP) and behaves like a nearly
"perfect" fluid \cite{RHIC-EXP,RHIC-THEO}. The AdS/CFT duality gives
a lower bound $\eta/s=1/4\pi$ \cite{bound}. Therefore, it is
conjectured that the sQGP created at RHIC might be the most perfect
fluid observed in nature.

\begin{figure}
\includegraphics[width=7cm]{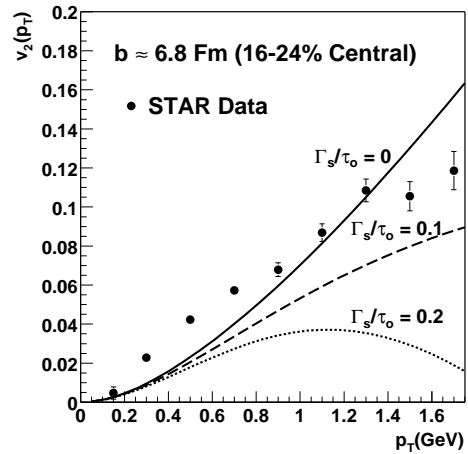}
\caption{Elliptic flow $v_2$ as a function of $p_T$ for different
values of $\Gamma_s/\tau_0$. The figure taken from
Ref.\cite{Hydro-Teaney}.} \label{fig-ShearV2}
\end{figure}

However, a perfect fluid should have both vanishing shear and bulk
viscosities.

The perturbative QCD calculation gives $\zeta/s=0.02 \alpha_s^2$ for
$0.06<\alpha_s<0.3$ \cite{Arnold-bulk}. In the hydrodynamic
simulation used to describe the evolution of the fireball created at
RHIC, the bulk viscosity $\zeta$ has often been neglected. The zero
bulk viscosity is for a conformal equation of state and also a
reasonable approximation for the weakly interacting gas of quarks
and gluons.  However, recent lattice QCD results show that the bulk
viscosity over entropy density ratio $\zeta/s$ rises dramatically up
to the order of $1.0$ near the critical temperature $T_c$
\cite{LAT-xis-KT,LAT-xis-KKT,LAT-xis-Meyer}. (There are still some
subtle issues to determine the bulk viscosity of QCD through
calculating the correlations of the energy-momentum tensor on the
lattice, see more detailed discussion in Ref.
\cite{correlation-Karsch}.) The sharp peak of bulk viscosity at
$T_c$ has also been observed in the linear sigma model
\cite{bulk-Paech-Pratt} and in the real scalar model
\cite{Li-Huang}. The increasing tendency of $\zeta/s$ has been shown
in a massless pion gas \cite{bulk-Chen} and in the NJL model below
$T_c$ \cite{Bulk-Sasaki}. The large bulk viscosity near phase
transition is related to the non-conformal equation of state
\cite{LAT-EOS-G, LAT-EOS-Nf2}, and the correlation between the bulk
viscosity and the conformal anomaly has been investigated in Ref.
\cite{Bulk-Nicola}.

\begin{figure}
\includegraphics[width=7cm]{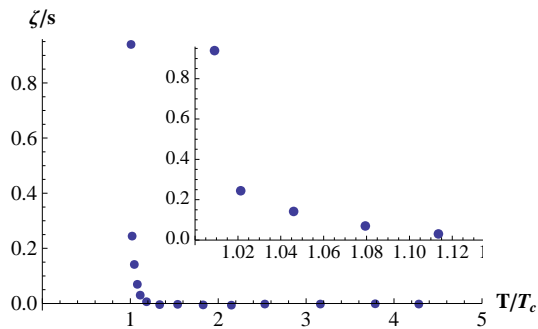}
\caption{The bulk viscosity over entropy density ratio as a function
of scaled temperature $T/T_c$. The figure is taken from
Ref.\cite{LAT-xis-KT}.} \label{fig-bulk}
\end{figure}

The sharp rise of the bulk viscosity will lead to the breakdown of
the hydrodynamic approximation around the critical temperature. The
effect of large bulk viscosity on hadronization and freeze-out
processes of QGP created at heavy ion collisions has been discussed
in Refs.
\cite{bulk-Mishustin,bulk-Muller,bulk-review-Kharzeev,bulk-kapusta}.
The authors of Ref. \cite{bulk-Mishustin} pointed out the
possibility that a sharp rise of bulk viscosity near phase
transition induces an instability in the hydrodynamic flow of the
plasma, and this mode will blow up and tear the system into
droplets. Another scenario is pointed out in Ref.
\cite{LAT-xis-KT,bulk-review-Kharzeev} that the large bulk viscosity
near phase transition might induce ``soft statistical
hadronization", i.e. the expansion of QCD matter close to the phase
transition is accompanied by the production of many soft partons,
which may be manifested through both a decrease of the average
transverse momentum of the resulting particles and an increase in
the total particle multiplicity.

\subsection{Searching for the critical end point}

\begin{figure}
\includegraphics[width=7cm]{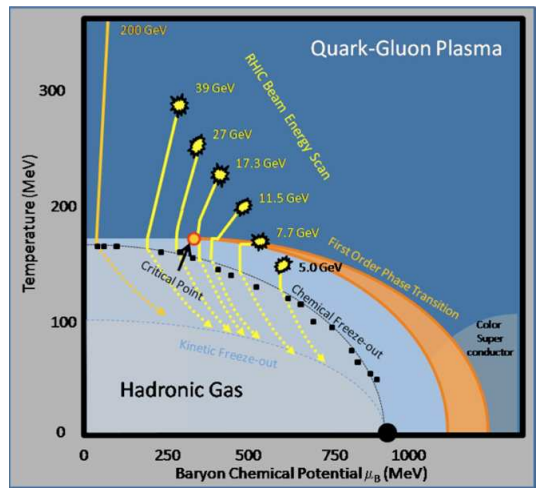}
\caption{Searching for the critical end point at RHIC.}
\label{fig-CEP}
\end{figure}

At small baryon chemical potential $\mu$, for QCD with two massless
quarks, the spontaneously broken chiral symmetry is restored at
finite temperature, and it is shown from lattice QCD
\cite{CEP-lattice} and effective QCD models \cite{CEP-models} that
this phase transition is of second order and belongs to the
universality class of $O(4)$ spin model in three dimensions
\cite{Pisarski-Wilczek}. For real QCD with two quarks of small mass,
the second order phase transition becomes a smooth crossover at
finite temperature. At finite baryon chemical potential, there are
still no reliable results from lattice QCD due to the severe fermion
sign problem. However QCD effective models \cite{CEP-models} suggest
that the chiral phase transition at finite $\mu$ is of first order.
It is expected that there exists a critical end point (CEP) in the
$T-\mu$ QCD phase diagram. The CEP is defined as the end point of
the first order phase transition, and belongs to the $Z(2)$ Ising
universality class \cite{Uni-CEP}. The signature of CEP has been
suggested in Refs. \cite{Sig-CEP}. The precise location of the CEP
is still unknown. In the future plan, RHIC is going to lower the
energy and trying to locate the CEP as shown in Fig. \ref{fig-CEP}.

\begin{figure}[h]
\includegraphics[angle=90,width=7cm]{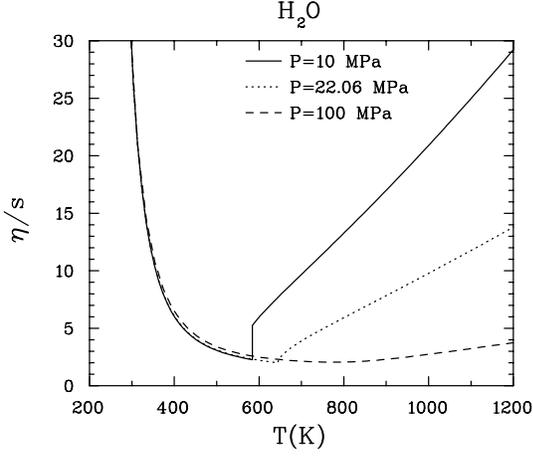}
\caption{The shear viscosity over entropy density ratio $\eta /s$ in
the water system. The figure is taken from
Ref.\cite{Csernai:2006zz}. } \label{fig-etaovers-water}
\end{figure}

Recently, the authors of Ref. \cite{Csernai:2006zz,etas-CEP}
suggested using the shear viscosity over entropy density ratio
$\eta/s$ to locate the CEP by observing the ratio of $\eta/s$
behaves differently in systems of water, helium and nitrogen in
first-, second-order phase transitions, see the system of water for
example in Fig. \ref{fig-etaovers-water}. The ratio of $\eta/s$
shows a cusp at $T_c$ for second order phase transition, and a
shallow valley near $T_c$ for cross-over, and shows a jump at $T_c$
for first-order phase transition.

Due to the complexity of QCD in the regime of strong coupling,
results on hot quark matter from lattice calculation and
hydrodynamic simulation are still lack of analytic understanding. In
recent years, the anti-de Sitter/conformal field theory (AdS/CFT)
correspondence has generated enormous interest in using thermal
${\cal N} = 4$ super-Yang-Mills theory (SYM) to understand sQGP. The
shear viscosity to entropy density ratio $\eta/s$ is as small as
$1/4\pi$ in the strongly coupled SYM plasma \cite{bound}. However, a
conspicuous shortcoming of this approach is the conformality of SYM:
the square of the speed of sound $c_s^2$ always equals to $1/3$ and
the bulk viscosity is always zero at all temperatures in this
theory. Though $\zeta/s$ at $T_c$ is non-zero for a class of black
hole solutions resembling the equation of state of QCD,  the
magnitude is less than $0.1$ \cite{Gubser-EOS}, which is too small
comparing with lattice QCD results.

An alternative nonperturbative approach to study QCD phase
transition is by using effective models. In the following, we
investigate the thermodynamical and transport properties in two toy
models, one is the simplest real scalar model
\cite{etas-scalar,Li-Huang}, the other is more relativistic QCD
effective model, i.e, the Polyakov-linear-sigma model (PLSM)
\cite{Mao:2009aq}, which can describe chiral phase transition as
well as deconfinement phase transition successfully.

\vskip 0.3cm
{\bf Real scalar model}
\vskip 0.3cm

We introduce the real scalar theory including the sextet interaction
which is described by the Lagrangian
\begin{equation}
\mathcal{L}=\frac{1}{2}(\partial _{\mu }\phi )^{2}-\frac{1}{2}a\phi ^{2}-%
\frac{1}{4}b\phi ^{4}-\frac{1}{6}c\phi ^{6} + H \phi.
\end{equation}%
When $H=0$, this theory is invariant under $\phi \rightarrow -\phi $
and has a $Z_{2}$ symmetry. here $a,b,c$ are model parameters, which
determine the vacuum properties. The system at finite temperature
will be evaluated in the Cornwall-Jackiw-Tomboulis (CJT) formalism
\cite{CJT}. We will discuss the following four cases: 1) $c=0, b>0,
a>0, H=0$, the system is always in
the symmetric phase. 2) $c=0, b>0, a<0, H=0$, the vacuum at $T=0$ breaks the $%
Z_{2}$ symmetry spontaneously, and the symmetry is restored at higher $%
T $ with a second-order phase transition. 3) $c=0, b>0, a<0,
H\neq0$, the $Z(2)$ symmetry is explicitly broken, and the system
will experience a crossover at high temperature. 4) $c>0,b<0, a>0,
H=0$, the broken symmetry is restored at high $T$ with a first-order
phase transition.

If symmetry is spontaneously broken in the vacuum, $\phi$ has a
vacuum expectation value ${\bar \phi}$, (in the case of no symmetry
breaking, ${\bar \phi}=0$ in the vacuum), we shift the field as
$\phi\rightarrow {\bar \phi}+ {\hat \phi}$. In terms of the shifted
field, the Lagrangian is given by
\begin{eqnarray}
\mathcal{L}&=&\mathcal{L}_0(\bar{\phi})+\frac{1}{2}(\partial _{\mu
}{\hat \phi} )^{2}- \frac{1}{2}m_0^2{\hat \phi}^2-(b{\bar
\phi}+\frac{10}{3}c{\bar \phi}^3){\hat \phi}^3 \nonumber
\\
&-&(\frac{b}{4}+\frac{5}{2}c{\bar \phi}^2){\hat \phi}^4-c{\bar
\phi}{\hat \phi}^5-\frac{1}{6}c{\hat \phi}^{6},
\end{eqnarray}%
with
\begin{eqnarray}
\mathcal{L}_0(\bar{\phi})=\frac{a}{2}~\bar{\phi}^{2}+\frac{b}{4}~\bar{\phi}^{4}+\frac{c}{6}~\bar{\phi}%
^{6} -H\bar{\phi}.
\end{eqnarray}
It is noticed that the new field ${\hat \phi}$ obtains a tree-level
mass of $m_{0}^{2}=a+3b~\bar{\phi}^{2}+5c~\bar{\phi}^{4}$. The
induced interaction terms including the cubic interaction term with
coupling strength $b{\bar \phi}+10/3 c{\bar \phi}^3$, the quartic
term with coupling strength $b/4+5/2c{\bar \phi}^2$, the quintic
term with coupling strength $c{\bar \phi}$, and the six-point
interaction term with coupling strength $1/6 c$.

Assuming translation invariance, we consider effective potential
$\Omega$ instead of effective action $\Gamma$, these two quantities
are related via:
\begin{equation}
\Gamma=-\frac{V}{T}\Omega,
\end{equation}
where $V$ is the 3-volume of the system. The effective potential in
the CJT formalism reads
\begin{eqnarray}
\Omega[\bar{\phi},\bar{G}] &=& \Omega_0(\bar{\phi})\,
+\,\Omega_{2}[\bar{\phi},\bar{G}] \nonumber
\\
& + & \frac{1}{2}\int_{K}\left[ \,\ln
\bar{G}^{-1}(K)+\bar{G}_{0}^{-1}(K)\,\bar{G}(K)-1\,\right]
 ,
\end{eqnarray}%
where $\Omega_0(\bar{\phi})=\mathcal{L}_0(\bar{\phi})$ is the
tree-level potential, and ${\bar G}({\bar G}_{0})$ is the
full(tree-level) propagator:
\begin{equation}
\bar{G}^{-1}(K,\bar{\phi})=-K^{2}+M^{2}(\bar{\phi})\;,\newline
\bar{G}_{0}^{-1}(K,\bar{\phi})=-K^{2}+m_{0}^{2}(\bar{\phi})\;.
\end{equation}%

In the Hartree approximation, the momentum dependent contributions
are neglected, $\Omega_2$ denotes the contribution from two-particle
irreducible diagrams, and takes the form of
\begin{equation}
\Omega_{2}[\bar{\phi},{\bar G}]=\left(
\frac{3}{4}b+\frac{15}{2}c\bar{\phi}^{2}\right) \left(\int_K {\bar
G}(K)\right) ^{2}+\frac{15}{6}c \left( \int_K {\bar G}(K)
\right)^{3}. \label{V2}
\end{equation}%

The self-consistent one- and two-point Green's functions satisfy
\begin{equation}
\left. \frac{\delta \Omega}{\delta \bar{\phi}}\right\vert
_{\bar{\phi}=\phi
,{\bar G}=G}\equiv 0\;,\;\;\;\left. \frac{\delta \Omega}{\delta {\bar G}}%
\right\vert _{\bar{\phi}=\phi,{\bar G}=G}\equiv 0\;\;\;\;.
\end{equation}%

All thermodynamical information of the system is contained in the
grand canonical potential $\Omega$, evaluated at the mean field
level. The entropy density $s$ is determined by taking the
derivative of effective potential with respect to the temperature,
i.e,
\begin{eqnarray} \label{entropy}
s=-\partial \Omega(\phi)/\partial T .
\end{eqnarray}
As the standard treatment in lattice calculation, we introduce the
normalized pressure density $p$ which is normalized to vanish at
$T=\mu=0$ and the energy density $\varepsilon$ as
\begin{eqnarray}
p=-\Omega, \,\, \varepsilon=-p+ T s.
\end{eqnarray}
The equation of state $p(\varepsilon)$ is an important input into
hydrodynamics. The square of the speed of sound $C_s^2$ is related
to $p/\varepsilon$ and has the form of
\begin{equation}
C_s^2=\frac{{\rm d}p}{{\rm d}\varepsilon}=\frac{s}{T {\rm d}s/{\rm
d}T}=\frac{s}{C_v},
\end{equation}
where
\begin{eqnarray}
C_v=\partial \varepsilon/\partial T,
\end{eqnarray}
is the specific heat. At the critical temperature, the entropy
density as well as the energy density change most quickly with
temperature, thus one expect that $C_s^2$ should have a minimum at
$T_c$.

The shear viscosity $\eta $ is calculated by using the Boltzmann
equation \cite{Jeon}. The two-particle elastic scattering amplitude,
which governs particle collisions in the Boltzmann equation, is%
\begin{equation}
i\mathcal{T}=\lambda _{4}+\lambda _{3}^{2}\left[ \frac{1}{s-m^{2}}+\frac{1}{%
t-m^{2}}+\frac{1}{u-m^{2}}\right] ,  \label{sca}
\end{equation}%
where $s,t$ and $u$ are Mandelstam variables, and $\lambda _{3}=6\phi _{0}(b+%
\frac{10c}{3}\phi _{0}^{2}+10c \int_{K}\,\bar{G}(K,\bar{\phi}))$ and
$\lambda _{4}=12(\frac{b}{2}+5c\phi _{0}^{2}+5c
\int_{K}\,\bar{G}(K,\bar{\phi}))$ are effective couplings.

\begin{figure}
\includegraphics[width=7.5cm]{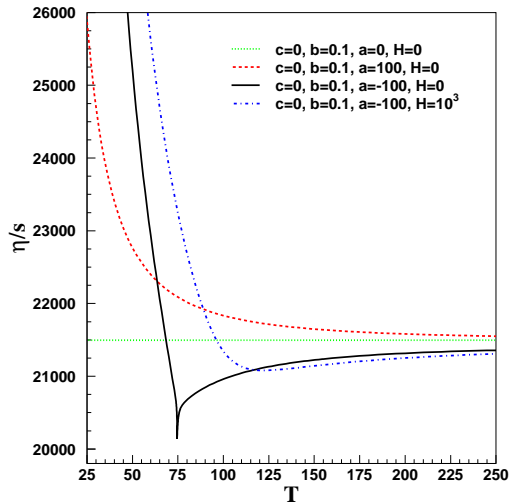}
\caption{The shear viscosity over entropy density $\eta/s$ as a
function of the temperature $T$, for cases with a second-order phase
transition\ (solid curve), a crossover (dash-dotted curve), and
with\ no phase transition for massive field (dashed curve) and
massless field (dotted curve). The figure is taken from
Ref.\cite{etas-scalar}.} \label{fig-etas-2nd-O2}
\end{figure}

\begin{figure}
\includegraphics[width=7.5cm]{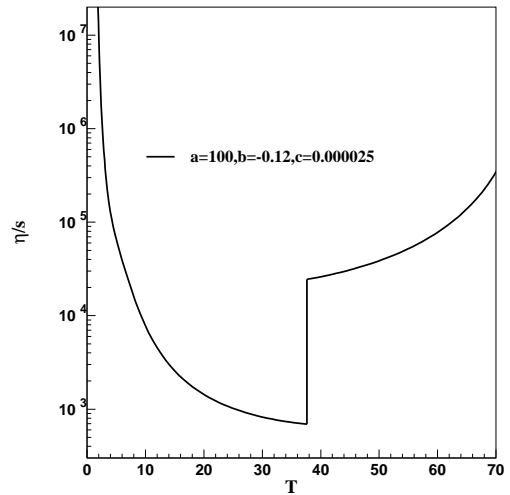}
\caption{The shear viscosity over entropy density $\eta/s$ as a
function of the temperature $T$, for the case of 1st order phase
transition.} \label{fig-etas-1st-O2}
\end{figure}

The shear viscosity over entropy density ratio $\eta/s$ in the real
scalar model in shown in Fig.\ref{fig-etas-2nd-O2} and
\ref{fig-etas-1st-O2} for different orders of phase transitions.
There is clearly a qualitative difference in the $\eta /s$ behavior
between cases with and without a phase transition. It is seen that
$\eta/s$ shows a cusp at $T_c$ for the case of 2nd-order phase
transition,  a shallow valley near $T_c$ for crossover, and shows a
jump at $T_c$ for the case of 1st-order phase transition. This
behavior is qualitatively the same as that in the classic systems
such as the in H$_{2}$O system as shown in Fig.
\ref{fig-etaovers-water}. If there is no phase transition. $\eta /s$
is always monotonically decreasing.

The bulk viscosity is related to the correlation function of the
trace of the energy-momentum tensor $\theta^\mu_\mu$:
\begin{equation}
\label{kubo} \zeta = \frac{1}{9}\lim_{\omega\to
0}\frac{1}{\omega}\int_0^\infty dt \int d^3r\,e^{i\omega t}\,\langle
[\theta^\mu_\mu(x),\theta^\mu_\mu(0)]\rangle \,.
\end{equation}
According to the result derived from low energy theorem, in the low
frequency region, the bulk viscosity takes the form of
\cite{LAT-xis-KT,LAT-xis-KKT}
\begin{eqnarray}\label{ze}
\,\zeta &=& \frac{1}{9\,\omega_0}\left\{ T^5\frac{\partial}{\partial
T}\frac{(\varepsilon-3p)}{T^4}
+16|\varepsilon_v|\right\}\,, \nonumber \\
 & = & \frac{1}{9\,\omega_0} \left\{- 16 \varepsilon+9 T S + T C_v + 16 |\varepsilon_v| \right\}\,.
\end{eqnarray}
with the negative vacuum energy density
$\varepsilon_v=\Omega_v=\Omega(\phi)|_{T=0}$, and the parameter
$\omega_0 = \omega_0(T)$ is a scale at which the perturbation theory
becomes valid. From the above formula, we can see that the bulk
viscosity is proportional to the specific heat $C_v$ near phase
transition, thus $\zeta/s$ behaves as $1/C_s^2$ near $T_c$ in this
approximation.

\begin{figure}
\includegraphics[width=7.5cm]{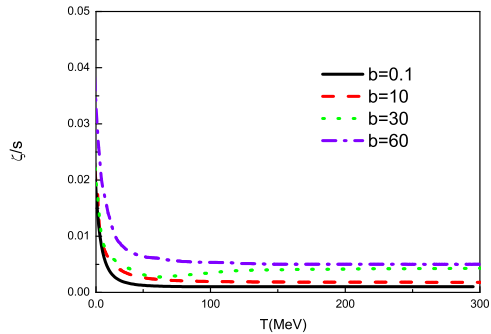}
\caption{The bulk viscosity over entropy density $\zeta/s$ as a
function of $T$ for the case without phase transition in the real
scalar model. The figure is taken from Ref.\cite{Li-Huang}.}
\label{fig-zetas-Symm}
\end{figure}

\begin{figure}
\includegraphics[width=7.5cm]{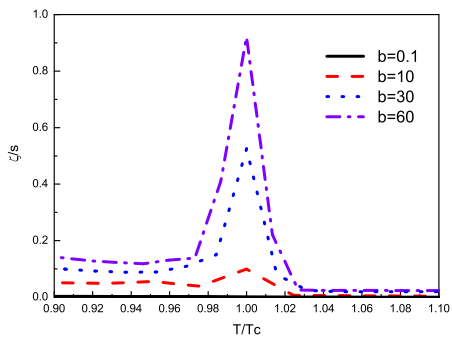}
\caption{The bulk viscosity over entropy density $\zeta/s$ as a
function of $T$ for a 2nd-order phase transition in the real scalar
model in the real scalar model. The figure is taken from
Ref.\cite{Li-Huang}.} \label{fig-zetas-2nd}
\end{figure}

\begin{figure}
\includegraphics[width=7.5cm]{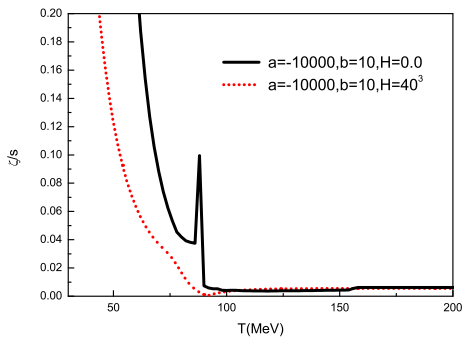}
\caption{The bulk viscosity over entropy density $\zeta/s$ as a
function of $T$ for the case of crossover (the solid line). The
figure is taken from Ref.\cite{Li-Huang}.} \label{fig-zetas-cross}
\end{figure}

\begin{figure}
\includegraphics[width=7.5cm]{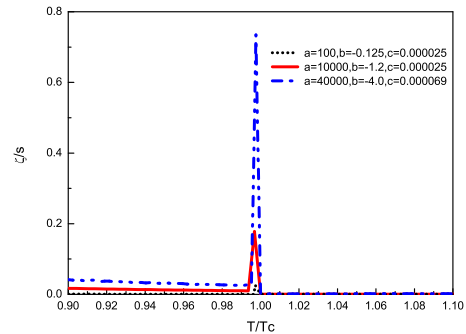}
\caption{The bulk viscosity over entropy density $\zeta/s$ as a
function of $T$ for a 1st-order phase transition in the real scalar
model. The figure is taken from Ref.\cite{Li-Huang}.}
\label{fig-zetas-1st}
\end{figure}

The bulk viscosity over entropy density ratio $\zeta/s$ as a
function of $T$ is shown in
Figs.\ref{fig-zetas-Symm},\ref{fig-zetas-2nd},\ref{fig-zetas-cross},
and \ref{fig-zetas-1st}. It is shown that in the case without
symmetry breaking, the bulk viscosity over entropy density $\zeta/s$
decreases monotonically with the increase of the temperature. In the
case of 2nd-order phase transition, $\zeta/s$ decreases with $T$ at
low temperature region, then rises up at the critical temperature
$T_c$ and shows an upward cusp, and decreases further in the
temperature $T>T_c$. In the case of crossover, it is observed the
cusp behavior of $\zeta/s$ is washed out. In the case of 1st-order
phase transition, $\zeta/s$ shows divergent behavior at $T_c$.

From Refs.\cite{etas-scalar, Csernai:2006zz}, we know that $\eta/s$
shows a shallow valley in the case of crossover and a jump at $T_c$
for first-order phase transition. But it is hard to distinguish
whether the system experiences a crossover or first-order phase
transition just from the value of $\eta/s$ extracted from the
elliptic flow $v_2$.

From our results in the real scalar model, it is found that the
ratio of $\zeta/s$ shows a very sharp peak at $T_c$ in the case of
first order phase transition, and there is no obvious change of
$\zeta/s$ for crossover. As pointed out in Ref.
\cite{bulk-Mishustin} that a sharp rise of bulk viscosity near phase
transition induces an instability in the hydrodynamic flow of the
plasma, and this mode will blow up and tear the system into
droplets. Therefore, one can distinguish whether the system
experiences a first order phase transition or a crossover from
observables at RHIC experiments. This result supports the idea of
using $\zeta/s$ to locate the CEP as suggested in Ref.
\cite{LAT-xis-KKT}.

\vskip 0.3cm {\bf The Polyakov-linear-sigma model}\vskip 0.3cm

We have shown the behavior of shear viscosity over entropy density
$\eta/s$ and bulk viscosity over entropy density $\zeta/s$ near
$T_c$ for different orders of phase transitions in a toy model, i.e,
the real scalar model. In the following, we use a more realistic QCD
effective model, i.e, the Polyakov-linear-sigma model (PLSM), which
is described by the Lagrangian \cite{Schaefer:2008ax}
\begin{eqnarray}\label{plsm}
\mathcal{L}=\mathcal{L}_{chiral}-\mathbf{\mathcal{U}}(\phi,\phi^*,T)
\end{eqnarray}
where we have separated the contribution of chiral degrees of
freedom and the Polyakov loop. The chiral part of the Lagrangian,
$\mathcal{L}_{chiral}= \mathcal{L}_q+\mathcal{L}_m$ consists of the
fermionic part
\begin{eqnarray}
\mathcal{L}_q=\sum_f \overline{\psi}_f(i\gamma^{\mu}
D_{\mu}-gT_a(\sigma_a+i \gamma_5 \pi_a))\psi_f \label{lfermion}
\end{eqnarray}
and the purely mesonic contribution
\begin{eqnarray}
\mathcal{L}_m &=&
\mathrm{Tr}(\partial_{\mu}\Phi^{\dag}\partial^{\mu}\Phi-m^2
\Phi^{\dag} \Phi)-\lambda_1 [\mathrm{Tr}(\Phi^{\dag} \Phi)]^2
\nonumber\\&& -\lambda_2 \mathrm{Tr}(\Phi^{\dag}
\Phi)^2+c[\mathrm{Det}(\Phi)+\mathrm{Det}(\Phi^{\dag})]
\nonumber\\&&+\mathrm{Tr}[H(\Phi+\Phi^{\dag})],\label{lmeson}
\end{eqnarray}
the sum is over the three flavors (f=1,2,3 for u, d, s). In the
above equation we have introduced a flavor-blind Yukawa coupling $g$
of the quarks to the mesons and the coupling of the quarks to a
background gauge field $A_{\mu}=\delta_{\mu 0}A_0$ via the covariant
derivative $D_{\mu}=\partial_{\mu}-i A_{\mu}$. The $\Phi$ is a
complex $3 \times 3$ matrix and is defined in terms of the scalar
$\sigma_a$ and pseudoscalar $\pi_a$ meson nonets,
\begin{eqnarray}
\Phi=T_a(\sigma_a+i\pi_a ).
\end{eqnarray}
The $3 \times 3$ matrix $H$ breaks the symmetry explicitly and is
chosen as
\begin{eqnarray}
H=T_a h_a,
\end{eqnarray}
where $h_a$ are nine external fields. The $T_a=\lambda_a/2$ are the
generators of the $U(3)$ symmetry, $\lambda_a$ are the Gell-Mann
matrices with $\lambda_0=\sqrt{\frac{2}{3}}\textbf{1}$. The $T_a$
are normalized to $\mathrm{Tr}(T_a T_b)=\delta_{ab}/2$ and obey the
$U(3)$ algebra with $[T_a,T_b]=i f_{abc}T_c$ and
$\{T_a,T_b\}=d_{abc} T_c$ respectively, here $f_{abc}$ and $d_{abc}$
for $a,b,c=1,...,8$ are the standard antisymmetric and symmetric
structure constants of $SU(3)$ group and
\begin{eqnarray}
f_{ab0}\equiv 0, \qquad d_{ab0}=\sqrt{\frac{2}{3}}\delta_{ab}.
\end{eqnarray}

The quantity $\mathbf{\mathcal{U}}(\phi,\phi^*,T)$ is the
Polyakov-loop effective potential expressed by the dynamics of the
traced Polyakov loop
\begin{eqnarray}
\phi=(\mathrm{Tr}_c L)/N_c, \qquad \phi^*=(\mathrm{Tr}_c
L^{\dag})/N_c.
\end{eqnarray}
The Polyakov loop $L$ is a matrix in color space and explicitly
given by
\begin{eqnarray}
L(\vec{x})=\mathcal{P}\mathrm{exp}\left[i\int_0^{\beta}d \tau
A_4(\vec{x},\tau)\right],
\end{eqnarray}
with $\beta=1/T$ being the inverse of temperature and $A_4=iA^0$. In
the Polyakov gauge, the Polyakov-loop matrix can be given as a
diagonal representation \cite{Fukushima:2003fw}. The coupling
between Polyakov loop and quarks is uniquely determined by the
covariant derivative $D_{\mu}$ in the PLSM Lagrangian in
Eq.(\ref{plsm}), and in the chiral limit, this Lagrangian is
invariant under the chiral flavor group, just like the original QCD
Lagrangian. The trace of the Polyakov-loop, $\phi$ and its conjugate
$\phi^*$ can be treated as classical field variables in this work.

The temperature dependent effective potential
$\mathbf{\mathcal{U}}(\phi,\phi^*,T)$ is used to reproduce the
thermodynamical behavior of the Polyakov loop for the pure gauge
case in accordance with lattice QCD data, and it has the $Z(3)$
center symmetry like the pure gauge QCD Lagrangian. In the absence
of quarks, we have $\phi=\phi^*$ and the Polyakov loop is taken as
an order parameter for deconfinement. For low temperatures,
$\mathbf{\mathcal{U}}$ has a single minimum at $\phi=0$, while at
high temperatures it develops a second one which turns into the
absolute minimum above a critical temperature $T_0$, and the $Z(3)$
center symmetry is spontaneously broken. In this paper, we will use
the potential $\mathbf{\mathcal{U}}(\phi,\phi^*,T)$ proposed in
Ref.\cite{Ratti:2005jh}, which has a polynomial expansion in $\phi$
and $\phi^*$:
\begin{eqnarray}
\frac{\mathbf{\mathcal{U}}(\phi,\phi^*,T)}{T^4}=-\frac{b_2(T)}{2}|\phi|^2-\frac{b_3
}{6}(\phi^3+\phi^{*3})+\frac{b_4}{4}(|\phi|^2)^2, \nonumber \\
\end{eqnarray}
with
\begin{eqnarray}
b_2(T)=a_0+a_1\left(\frac{T_0}{T}\right)+a_2\left(\frac{T_0}{T}\right)^2+a_3\left(\frac{T_0}{T}\right)^3.
\end{eqnarray}
A precision fit of the constants $a_i,b_i$ is performed to reproduce
the lattice data for pure gauge theory thermodynamics and the
behavior of the Polyakov loop as a function of temperature. The
corresponding parameters are
\begin{eqnarray}
a_0=6.75,\qquad a_1=-1.95,\qquad a_2=2.625, \nonumber \\
 a_3=-7.44,\qquad
b_3=0.75,\qquad b_4=7.5.
\end{eqnarray}
The critical temperature $T_0$ for deconfinement in the pure gauge
sector is fixed at $270$ MeV, in agreement with the lattice results.

we obtain the thermodynamical potential density as
\begin{eqnarray}\label{potential}
\Omega(T,\mu_f)& = & \frac{-T \mathrm{ln}
\mathcal{Z}}{V}=U(\sigma_x,\sigma_y)+\mathbf{\mathcal{U}}(\phi,\phi^*,T)+\Omega_{\bar{\psi}
\psi}, \nonumber \\
\end{eqnarray}
with the quarks and antiquarks contribution
\begin{eqnarray}
& &\Omega_{\bar{\psi} \psi} = -2 T N_q \int \frac{d^3\vec{p}}{(2
\pi)^3} \{ \nonumber \\
& & \mathrm{ln} [1+3(\phi+\phi^* e^{-(E_q-\mu)/T})
e^{-(E_q-\mu)/T}+e^{-3 (E_q-\mu)/T}] \nonumber \\
& &  +\mathrm{ln} [ 1+3(\phi^*+\phi e^{-(E_q+\mu)/T})
e^{-(E_q+\mu)/T}+e^{-3 (E_q+\mu)/T}] \} \nonumber \\
& & -2 T N_s \int \frac{d^3\vec{p}}{(2 \pi)^3} \{ \nonumber
\\
& & \mathrm{ln} [ 1+3(\phi+\phi^* e^{-(E_s-\mu)/T})
e^{-(E_s-\mu)/T}+e^{-3 (E_s-\mu)/T}] \nonumber \\
& & +\mathrm{ln} [ 1+3(\phi^*+\phi e^{-(E_s+\mu)/T})
e^{-(E_s+\mu)/T}+e^{-3 (E_s+\mu)/T}] \}. \nonumber \\
\end{eqnarray}
Here, $N_q=2$, $N_s=1$, and $E_q=\sqrt{\vec{p}^2+m_q^2}$ is the
valence quark and antiquark energy for $u$ and $d$ quarks, for
strange quark $s$, it is $E_s=\sqrt{\vec{p}^2+m_s^2}$, and $m_q$,
$m_s$ is the constituent quark mass for $u$, $d$ and $s$. The purely
mesonic potential is
\begin{eqnarray}
U(\sigma_x,\sigma_y)&=&\frac{m^2}{2} (\sigma^2_x+\sigma^2_y)-h_x
\sigma_x-h_y \sigma_y-\frac{c}{2\sqrt{2}} \sigma^2_x \sigma_y
\nonumber \\
&+ & \frac{\lambda_1}{2} \sigma^2_x \sigma^2_y +\frac{1}{8} (2
\lambda_1 +\lambda_2)\sigma^4_x + \frac{1}{4}
(\lambda_1+\lambda_2)\sigma^4_y. \nonumber \\
\end{eqnarray}
Minimizing the thermodynamical potential in Eq.(\ref{potential})
with respective to $\sigma_x$, $\sigma_y$, $\phi$ and $\phi^*$, we
obtain a set of equations of motion
\begin{eqnarray}
\frac{\partial \Omega}{\partial \sigma_x}=0, \qquad \frac{\partial
\Omega}{\partial \sigma_y}=0, \qquad \frac{\partial \Omega}{\partial
\phi}=0, \qquad \frac{\partial \Omega}{\partial \phi^*}=0.
\end{eqnarray}
The set of equations can be solved for the fields as functions of
temperature $T$ and chemical potential $\mu$, and the solutions of
these coupled equations determine the behavior of the chiral order
parameter $\sigma_x$, $\sigma_y$ and the Polyakov loop expectation
values $\phi$, $\phi^*$ as a function of $T$ and $\mu$.

\begin{figure}
\includegraphics[width=7cm]{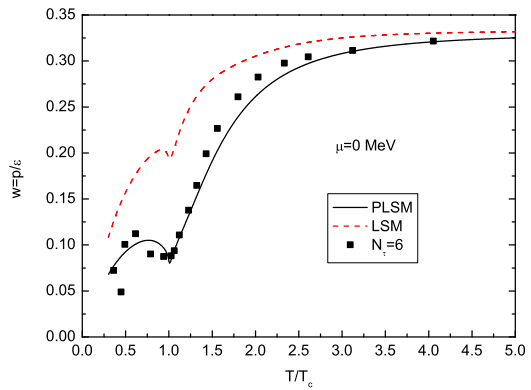}
\caption{The equation-of-state parameter $w(T)=p(T)/\varepsilon(T)$
for $\mu=0$ MeV. The Polyakov linear sigma model prediction (solid
line) and the linear sigma model prediction (dash line) are compared
with $N_f=2+1$ lattice QCD data for $N_{\tau}=6$. Lattice data taken
from Ref.\cite{LAT-EOS-Nf2}. The figure is taken from
Ref.\cite{Mao:2009aq}.} \label{fig-povere-PLSM}
\end{figure}

Fig.\ref{fig-povere-PLSM} shows the pressure density over energy
density $p/\varepsilon$, which is represented in terms of
equation-of-state (EOS) parameter, at zero density and finite
density, respectively. We observe that the pressure density over
energy density increases with temperature and saturates at high
temperature. Both the linear sigma model and the Polyakov linear
sigma model give very similar results at high temperature, the
pressure density over energy density $p/\varepsilon$ saturates at a
value smaller than $1/3$. Another common feature of the
$p/\varepsilon$ in the linear sigma model and the Polyakov linear
sigma model is that there is a bump appearing at low temperature
region, which is also observed in the lattice result. Around the
critical temperature $T_c$, the pressure density over energy density
$p/\varepsilon$ shows a downward cusp. However, the minimum value of
the $p/\varepsilon$ around $T_c$ is $0.2$ in the linear sigma model,
which is much larger than the result from the Polyakov linear sigma
model and the lattice QCD data. For the Polyakov linear sigma model,
the minimum of $p/\varepsilon$ around $T_c$ is $0.075$, which is
consistent with the lattice QCD data \cite{LAT-EOS-Nf2}.

\begin{figure}
\includegraphics[width=7cm]{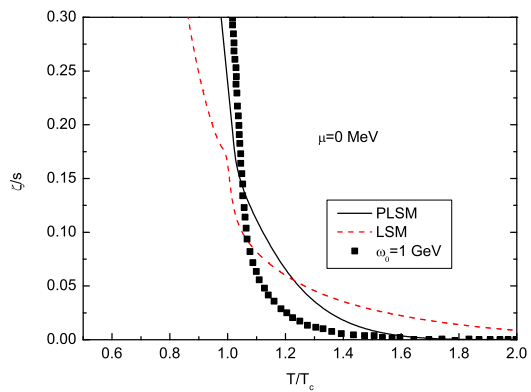}
\caption{The bulk viscosity over entropy density ratio $\zeta /s$ as
a function of the temperature for $\mu=0$ MeV. The solid line
denotes the Polyakov linear sigma model prediction and the dashed
line denotes the linear sigma model prediction. Lattice data taken
from Ref.\cite{LAT-xis-KKT}. The figure is taken from
Ref.\cite{Mao:2009aq}.} \label{fig-zetaovers-PLSM-T0}
\end{figure}

In Fig.\ref{fig-zetaovers-PLSM-T0}, we plot the bulk viscosity over
entropy density ratio $\zeta /s$ as a function of the temperature
for zero chemical potential. It is shown that, at zero chemical
potential $\mu=0$, the bulk viscosity over entropy density $\zeta
/s$ decreases monotonically with the increase of the temperature in
both the Polyakov linear sigma model and linear sigma model, and at
high temperature, $\zeta /s$ reaches its conformal value $0$. In
\cite{LAT-xis-KKT}, the bulk viscosity over entropy density of the
three flavor system is extracted from lattice result, which is shown
by the square. It is observed that $\zeta /s$ in PLSM near phase
transition is in very good agreement with the lattice result in
\cite{LAT-xis-KKT}, i.e, it rises sharply near phase transition.

\begin{figure}
\includegraphics[width=7cm]{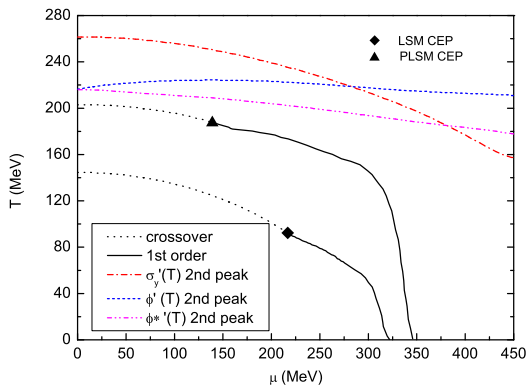}
\caption{The $T-\mu$ phase diagram in the Polyakov linear sigma
model. The figure is taken from Ref.\cite{Mao:2009aq}. }
\label{fig-CEP-PLSM}
\end{figure}

In Ref. \cite{Li-Huang}, we have investigated the equation of state
and bulk viscosity in the real scalar model and $O(4)$ model in the
case of 2nd order phase transition, crossover and 1st order phase
transition, and we have found that the thermodynamic properties and
transport properties in these simple models near the critical
temperature $T_c$ at strong coupling are similar to those of the
complex QCD system. In a more realistic QCD effective model, i.e,
the Polyakov linear sigma model \cite{Mao:2009aq}, we have
systematically investigated the thermodynamic properties and bulk
viscosity and found these properties match with lattice data very
well in the case of zero chemical potential. We further evaluate the
chiral phase transitions of $u,d$ and $s$ quarks and deconfinement
phase transition at finite temperature and finite density, and show
the $T-\mu$ phase structure of the Polyakov linear sigma model in
Fig.\ref{fig-CEP-PLSM}.

For the Polyakov linear sigma model, the result shows that the
critical end point is around $(T_E,\mu_E)=(188 ~{\rm MeV},
139.5~{\rm MeV})$, which is close to the lattice result $(T_E,
\mu_E)=(162\pm 2~{\rm MeV}, \mu_E=120 \pm 13~{\rm MeV})$
\cite{Fodor:2004nz}. For the linear sigma model without the Polyakov
loop, the critical end point is located at $(T_E, \mu_E) \simeq
(92.5 {\rm MeV}, 216~{\rm MeV})$. The critical chemical potential
$\mu_E$ in PLSM is much lower than that in the PNJL model with three
quark flavors where the predicted critical end point is $\mu_E >
300$ MeV\cite{Fu:2007xc,Ciminale:2007sr}.

The chiral phase transition for the strange quark and the
deconfinement phase transition in the $T-\mu$ plane are shown by the
dash-dotted line and dotted line, respectively. It is found that
with the increase of chemical potential, the critical temperature
for strange quark to restore chiral symmetry decreases. However, for
the deconfinement phase transition, with the increase of chemical
potential, the deconfinement critical temperature keeps almost a
constant around $220$ MeV. It can be seen that in the Polyakov
linear sigma model, there exists two-flavor quarkyonic phase
\cite{McLerran:2007qj} at low density, where the $u,d$ quarks
restore chiral symmetry but still in confinement, and three-flavor
quarkyonic phase at high density, where the $u,d,s$ quarks restore
chiral symmetry but still in confinement.

\begin{figure}
\includegraphics[width=7cm]{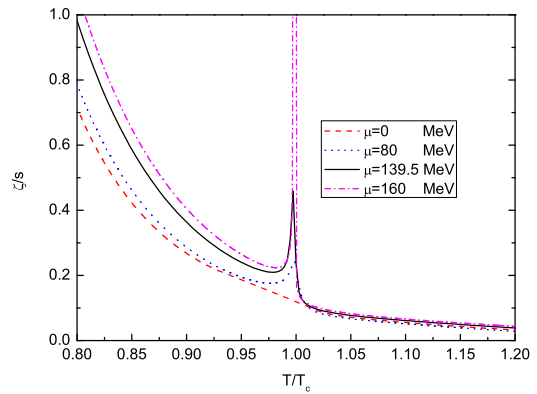}
\caption{The bulk viscosity over entropy density ratio $\zeta /s$ in
the PLSM for different chemical potentials as functions of $T/T_c$.
The figure is taken from Ref.\cite{Mao:2009aq}. }
\label{fig-zetaovers-PLSM-CEP}
\end{figure}

Because the Polyakov-loop in the PLSM is not introduced dynamically,
it is difficult to calculate the transport properties from the
Boltzmann equation. However, we can use Eq.(\ref{ze}) to calculate
the bulk viscosity. In Fig.\ref{fig-zetaovers-PLSM-CEP}, we plot the
bulk viscosity over entropy density ratio $\zeta /s$ as a function
of the temperature for different chemical potentials. It shows
$\zeta /s$ as function of the scaled temperature $T/T_c$ for
different chemical potentials with $\mu=0,80,139.5, 160$ MeV. We can
see that when the chemical potential increases up to $\mu=80$ MeV,
there is an upward cusp appearing in $\zeta /s$ right at the
critical temperature $T_c$. With the increase of the chemical
potential, the upward cusp becomes sharper, and the height of the
cusp increases. At the critical end point $\mu_E$ and when
$\mu>\mu_E$ for the first order phase transition, $\zeta /s$ becomes
divergent at the critical temperature.

As discussed earlier, the sharp rise of the bulk viscosity will lead
to the breakdown of the hydrodynamic approximation around the
critical temperature, and will affect the hadronization and
freeze-out processes of QGP created at heavy ion collisions Refs.
\cite{bulk-Mishustin,bulk-Muller,bulk-review-Kharzeev,bulk-kapusta}.
For example, in Ref. \cite{bulk-Mishustin}, it is pointed out that a
sharp rise of bulk viscosity near phase transition might induce an
instability in the hydrodynamic flow of the plasma, and this mode
will blow up and tear the system into droplets. Another scenario is
pointed out in Ref. \cite{LAT-xis-KT,bulk-review-Kharzeev} that the
large bulk viscosity near phase transition might induce ``soft
statistical hadronization", i.e. the expansion of QCD matter close
to the phase transition is accompanied by the production of many
soft partons, which may be manifested through both a decrease of the
average transverse momentum of the resulting particles and an
increase in the total particle multiplicity. Therefore the critical
end point might be located through the observables which are
sensitive to the ratio of bulk viscosity over entropy density.

However, it is noticed that the results of bulk viscosity in this
paper are based on Eq. (\ref{ze}), where the ansatz for the spectral
function \begin{equation}
\frac{\rho(\omega,\vec{0})}{\omega}=\frac{9\zeta}{\pi}
\frac{\omega_0^2}{\pi (\omega^2+\omega^2)} \label{ansatz}
\end{equation}
has been used in the small frequency, and $\omega_0$ is a scale at
which the perturbation theory becomes valid. In our calculation,
$\omega_0=1~{\rm GeV}$, its magnitude at $T_c$ is in agreement with
that obtained in ChPT for massive pion gas system in Ref.
\cite{Bulk-Nicola}. Qualitatively, the bulk viscosity corresponds to
nonconformality, thus it is reasonable to observe a sharp rising of
bulk viscosity near phase transition.  Ref. \cite{Bulk-Nicola} has
investigated the correlation between the bulk viscosity and
conformal breaking, and supports the results in
Ref.\cite{LAT-xis-KT,LAT-xis-KKT}. The sharp rising of bulk
viscosity has also been observed by another lattice result
\cite{LAT-xis-Meyer} and in the linear sigma model
\cite{bulk-Paech-Pratt}. However, till now, no full calculation has
been done for the bulk viscosity. The frequency dependence of the
spectral density has been analyzed in Refs. \cite{Moore} and
\cite{correlation-Karsch} and the limitation of the ansatz
Eq.(\ref{ansatz}) has been discussed. From Eq. (\ref{ze}), we see
that the bulk viscosity is dominated by $C_v$ at $T_c$. If $C_v$
diverges at $T_c$, the bulk viscosity should also be divergent at
the critical point and behave as $t^{-\alpha}$. However, the
detailed analysis in the Ising model in Ref. \cite{Onuki} shows a
very different divergent behavior $\zeta \sim t^{-z\nu +\alpha}$,
with $z\simeq 3 $ the dynamic critical exponent and $\nu\simeq
0.630$ the critical exponent in the Ising system. More careful
calculation on the bulk viscosity is needed in the future.

\section{Color superconducting phases}
\label{sec-CSC}

It is known that the deconfined cold dense quark matter is in color
superconducting phase.

Let us start with the system of free fermion gas. Fermions obey the
Pauli exclusion principle, which means no two identical fermions can
occupy the same quantum state. The energy distribution for fermions
(with mass $m$) has the form of
\begin{eqnarray}
f(E_p)=\frac{1}{e^{\beta(E_p-\mu)}+1}, ~~ \beta=1/T,
\end{eqnarray}
here $E_p=\sqrt{p^2+m^2}$, $\mu$ is the chemical potential and $T$
is the temperature. At zero temperature, $f(E_p)=\theta(\mu-E_p)$.
The ground state of the free fermion gas is a filled Fermi sea,
i.e., all states with the momenta less than the Fermi momentum
$p_F=\sqrt{\mu^2-m^2}$ are occupied, and the states with the momenta
greater than the Fermi momentum $p_F$ are empty. Adding or removing
a single fermion costs no free energy at the Fermi surface.

For the degenerate Fermi gas, the only relevant fermion degrees of
freedom are those near the Fermi surface. Considering two fermions
near the Fermi surface, if there is a net attraction between them,
it turns out that they can form a bound state, i.e., Cooper pair
\cite{Cooper}. The binding energy of the Cooper pair $\Delta(K)$
($K$ the total momentum of the pair), is very sensitive to $K$,
being a maxium where $K=0$. There is an infinite degeneracy among
pairs of fermions with equal and opposite momenta at the Fermi
surface. Because Cooper pairs are composite bosons, they will occupy
the same lowest energy quantum state at zero temperature and produce
a Bose-Einstein condensation. Thus the ground state of the Fermi
system with a weak attractive interaction is a complicated coherent
state of particle-particle Cooper pairs near the Fermi surface
\cite{BCS}. Exciting a quasiparticle and a hole which interact with
the condensate requires at least the energy of $2 \Delta$.

In QED case in condensed matter, the interaction between two electrons by
exchanging a photon is repulsive. The attractive interaction to form
electron-electron Cooper pairs is by exchanging a phonon, which is a collective
excitation of the positive ion background. The Cooper pairing of the electrons
breaks the electromagnetic gauge symmetry, and the photon obtains an effective
mass. This indicates the Meissner effect \cite{Meissner}, i.e., a superconductor
expels the magnetic fields.

In QCD case at asymptotically high baryon density, the
dominant interaction between two quarks is due to the one-gluon exchange.
This naturally provides an attractive interaction between two quarks.
The scattering amplitude for single-gluon exchange in an $SU(N_c)$ gauge theory
is proportional to
\begin{eqnarray}
(T_a)_{ki}(T_a)_{lj}&=&-\frac{N_c+1}{4N_c} (\delta_{jk}\delta_{il} -
\delta_{ik}\delta_{jl}) \\ \nonumber &+& \frac{N_c-1}{4N_c}
(\delta_{jk}\delta_{il} + \delta_{ik}\delta_{jl}). \label{T-OGE}
\end{eqnarray}
Where $T_a$ is the generator of the gauge group, and $i,j$ and $k,l$ are the fundamental
colors of the two quarks in the incoming and outgoing channels, respectively. Under
the exchange of the color indices of either the incoming or the outgoing quarks,
the first term is antisymmetric, while the second term is symmetric. For $N_c=3$,
Eq.~(\ref{T-OGE}) represents that the tensor product of two fundamental colors
decomposes into an (antisymmetric) color antitriplet and a (symmetric) color sextet,
\begin{eqnarray}
[{\bf 3}]^c \otimes [{\bf 3}]^c = [\bar{\bf 3}]_a^c \oplus [{\bf 6}]_s^c.
\end{eqnarray}
In Eq.~(\ref{T-OGE}), the minus sign in front of the antisymmetric
contribution indicates that the interaction in this antitriplet
channel is attractive, while the interaction in the symmetric sextet
channel is repulsive.

For cold dense quark matter, the attractive interaction in the color
antitriplet channel induces the condensate of the quark-quark Cooper pairs, and
the ground state is called the ``color superconductivity". Since the diquark cannot
be color singlet, the diquark condensate breaks the local color $SU(3)_c$ symmetry,
and the gauge bosons connected with the broken generators obtain masses.
Comparing with the Higgs mechanism of dynamical gauge symmetry breaking in the
Standard Model, here the diquark Cooper pair can be regarded as a composite Higgs
particle. The calculation of the energy gap and the critical temperature from the
first principles has been derived systematically in Refs.
\cite{weak-Son,weak-Schafer,weak-Dirk,weak-Tc-Dirk,weak-Hong,weak-HongShovkovy,weak-Ren,weak-Manuel}.

In reality, we are more interested in cold dense quark matter at moderate baryon density
regime, i.e., $\mu_q \sim 500 MeV$, which may exist in the interior of neutron stars.
It is likely that cold dense quark droplet might be created in the laboratory
through heavy ion collisions in GSI-SPS energy scale.
At these densities, an extrapolation of the asymptotic arguments becomes unreliable, we
have to rely on effective models. Calculations in the framework of pointlike
four-fermion interactions based on the instanton vertex
\cite{CS-1997,Instanton-Schafer,Berges-Rajagopal,Carter-Diakonov},
as well as in the Nambu--Jona-Lasinio (NJL) model
\cite{NJL-Klevansky,NJL-Aichelin,NJL-Buballa,NJL-Ebert,NJL-Huang} show that color
superconductivity does occur at moderate densities, and the magnitude of diquark gap
is around $100~{\rm MeV}$.

\subsection{Different color superconducting phases}
\label{csc-phases}

Even though the antisymmetry in the attractive channel signifies
that only quarks with different colors can form Cooper pairs, color
superconductivity has very rich phase structure because of its
flavor, spin and other degrees of freedom. In the following, I list
some of the known color superconducting phases.

\vskip 0.3cm
{\bf The 2SC phase}
\vskip 0.3cm

Firstly we consider a system with only massless $u$ and $d$ quarks,
assuming that the strange quark is much heavier than the up and down quarks. The color
superconducting phase with only two flavors is normally called the 2SC phase.

Renormalization group arguments \cite{weak-Son,RG-Schafer,RG-Evans} suggest that
possible quark pairs always condense in the $s-$wave. This means that the spin wave
function of
the pair is anti-symmetric. Since the diquark condenses in the color antitriplet
$\bar{\bf 3}_c$ channel, the color wave function of the pair is also
anti-symmetric. The Pauli principle requires that the total wave function
of the Cooper pair has to be antisymmetric under the exchange of the two
quarks forming the pair. Thus the flavor wave function has to be
anti-symmetric, too. This determines the structure of the order parameter
\begin{eqnarray}
\label{order2sc}
\Delta_{ij}^{\alpha\beta}=\Delta \epsilon_{ij}\epsilon^{\alpha\beta b},
\end{eqnarray}
where color indices $\alpha, \beta \in (r,g,b)$ and flavor indices
$i, j \in(u,d)$. From the order parameter Eq. (\ref{order2sc}), we can see
that the condensate picks a color direction (here the $blue$ direction, which is
arbitrarily selected). The ground state is invariant under
an $SU(2)_c$ subgroup of the color rotations that mixes the red and green colors, but
the blue quarks are singled out as different. Thus the color $SU(3)_c$ is broken down
to its subgroup $SU(2)_c$, and five of the gluons obtain masses, which indicates the
Meissner effect \cite{Meissner2SC}.

In the 2SC phase, the Cooper pairs are $ud-du$ singlets and the global flavor
symmetry $SU(2)_L \otimes SU(2)_R$ is intact, i.e., the chiral symmetry is not broken.
There is also an unbroken global symmetry which plays the role of $U(1)_B$. Thus
no global symmetry are broken in the 2SC phase.

\vskip 0.3cm
{\bf The CFL phase}
\vskip 0.3cm

In the case when the chemical potential is much larger than the strange
quark mass, we can assume $m_u=m_d=m_s=0$, and there are three degenerate massless
flavors in the system. The spin-0 order parameter should be color and flavor
anti-symmetric, which has the form of
\begin{eqnarray}
\Delta_{ij}^{\alpha\beta}= \Delta \sum_I \epsilon_{ijI} \epsilon^{\alpha\beta I},
\label{CFL-gap}
\end{eqnarray}
where color indices $\alpha, \beta \in (r,g,b)$ and flavor indices $i, j \in(u,d,s)$.
Writing $\sum_I\epsilon_{ijI} \epsilon^{\alpha\beta I}=\delta_i^{\alpha} \delta_j^{\beta}
-\delta_j^{\alpha} \delta_i^{\beta} $, we can see that the order parameter
\begin{eqnarray}
\Delta_{ij}^{\alpha\beta}= \Delta (\delta_i^{\alpha} \delta_j^{\beta}
-\delta_j^{\alpha} \delta_i^{\beta})
\end{eqnarray}
describes the color-flavor locked (CFL) phase proposed in Ref.~\cite{CFL-ARW}. Many
other different treatments
\cite{CFL-Chiral-Schaefer,CFL-Evans,CFL-Shovkovy}
agreed that a condensate of the form (\ref{CFL-gap}) is the dominant condensate in
three-flavor QCD.

In the CFL phase, all quark colors and flavors participate in the pairing.
The color gauge group is completely broken, and all eight gluons become
massive \cite{CFL-ARW,Meissner-CFL}, which ensures that there are no infrared
divergences associated with gluon propagators. Electromagnetism is no
longer a separate symmetry, but corresponds to gauging one of the flavor
generators. A rotated electromagnetism (``$\tilde{Q}$") remains unbroken.

Two global symmetries, the chiral symmetry and the baryon number, are broken in
the CFL phase, too. In zero-density QCD, the spontaneous breaking of chiral symmetry
is due to the condensation of left-handed quarks with right-handed quarks.
Here, at high baryon density, the chiral symmetry breaking occurs due to a
rather different mechanism: locking of the flavor rotations to color.
In the CFL phase, there is only pairing of left-handed quarks with left-handed
quarks, and right-handed quarks with right-handed quarks, i.e.,
\begin{eqnarray}
< \psi_{Li}^{\alpha}\psi_{Lj}^{\beta}>
= - < \psi_{Ri}^{\alpha}\psi_{Rj}^{\beta}>.
\end{eqnarray}
Where $L,R$ indicate left- and right-handed, respectively,
$\alpha,\beta$ are color indices and $i,j$ are flavor indices. A
gauge invariant form \cite{CFL-effective,CFL-Son}
\begin{eqnarray}
< \psi_{Li}^{\alpha}\psi_{Lj}^{\beta}
   {\bar \psi}_{R\alpha}^k {\bar \psi}_{R\beta}^l>
&\sim&  < \psi_{Li}^{\alpha}\psi_{Lj}^{\beta}>
   <{\bar \psi}_{R\alpha}^k {\bar \psi}_{R\beta}^l> \nonumber \\
&\sim& \Delta^2 \epsilon_{ijm}\epsilon^{klm}
\end{eqnarray}
captures the chiral symmetry breaking. The spectrum of excitations in the CFL phase
contains an octet of Goldstone bosons associated with the chiral symmetry breaking.
This looks remarkably like those at low density. In the excitation spectrum
of the CFL phase, there is another singlet $U(1)$ Goldstone boson related to the
baryon number symmetry breaking, which can be described using the order parameter
\begin{eqnarray}
<udsuds>\sim<\Lambda \Lambda>.
\end{eqnarray}
In QCD with three degenerate light flavors, the spectrum in the CFL phase looks
similar to that in the hyper-nuclear phase at
low-density. It is suggested that the low density hyper-nuclear phase and the high
density quark phase might be continuously connected \cite{Continuity-Schafer}.

\vskip 0.3cm
{\bf Spin-1 color superconductivity}
\vskip 0.3cm
In the case of only one-flavor quark system, due to the antisymmetry in
the color space, the Pauli principle requires that the Cooper pair has to
occur in a symmetric spin channel.
Therefore, in the simplest case, the Cooper pairs carry total spin one.
Spin-1 color
superconductivity was firstly studied in Ref.~\cite{Bailin-Love}, for more
recent and detailed
discussions about the spin-1 gap, its critical temperature and Meissner
effect, see
Refs.~ \cite{weak-Dirk,weak-Tc-Dirk,Spin1-Schafer,Spin1-Buballa,Spin1-Tc-Schmitt,Spin1-Meissner}.
For a review, see Ref. ~\cite{Schmitt-PhD}.

\vskip 0.3cm
{\bf Pairing with mismatch: LOFF, CFL-K, g2SC and gCFL}
\vskip 0.3 cm

To form the Cooper pair, the ideal case is when the two pairing quarks have the same
Fermi momenta, i.e., $p_{F,i}=p_{F,j}$ with $p_{F,i}=\sqrt{\mu_{F,i}^2-m_i^2}$, like
in the ideal 2SC, CFL, and spin-1 color superconducting phases. However, in reality,
the nonzero strange quark mass or the requirement of charge neutrality induces a
mismatch between the Fermi momenta of the two pairing quarks. When the mismatch is
very small, it has little effect on the Cooper pairing. While if the mismatch is very
large, the Cooper pair will be destroyed. The most interesting situation happens when
the mismatch is neither very small nor very large.

{\bf LOFF:} In the regime just on the edge of decoupling of the two pairing quarks
(due to the nonzero strange quark mass for the $qs$ Cooper pair with $q\in (u,d)$ or
the chemical potential difference for the $ud$ Cooper pair), a ``LOFF"
(Larkin-Ovchinnikov-Fulde-Ferrell) state may be formed. The LOFF state was firstly
investigated in the context of electron superconductivity in the presence of
magnetic impurities \cite{LO,FF}. It was found that near the unpairing transition,
it is favorable to form a state in which the Cooper pairs have nonzero momentum.
This is favored because it gives rise to a regime of phase space where each of the
two quarks in a pair can be close to its Fermi surface, and such pairs can be
created at low cost in free energy. This sort of condensates spontaneously break
translational and rotational invariance, leading to gaps which vary periodically
in a crystalline pattern. The crystalline color superconductivity has been
investigated in a series of papers, e.g., see
Refs. ~\cite{LOFF-1,LOFF-2,LOFF-3,LOFF-4,LOFF-5,LOFF-Bowers,LOFF-Ren,LOFF-Nardulli}.

{\bf CFL-K:} The strange quark mass $m_s$ induces an effective chemical potential
$\mu_s=m_s^2/(2p_F)$, and the effects of the strange quark mass can be quite dramatic.
In the CFL phase, the $K^+$ and $K^0$ modes may be unstable for large values of the
strange quark mass to form a kaon condensation
\cite{CFLK-Schafer,CFLK-Reddy,Schafer-Goldstone,Igor-Goldstone}.
In the framework of effective theory
\cite{CFL-effective,CFL-Son,CFL-mesons-Rho,CFL-mesons-Hong,CFL-mesons-Manuel},
the masses of the Goldstone bosons can be determined as
\begin{eqnarray}
\label{mgb}
 m_{\pi^\pm} &=& \mp\frac{m_d^2-m_u^2}{2p_F}+\left [
       \frac{4A}{f_{\pi}^2}(m_u+m_d)m_s\right ]^{1/2}, \nonumber \\
 m_{K^\pm} &=& \mp\frac{m_s^2-m_u^2}{2p_F}+\left [
       \frac{4A}{f_{\pi}^2}(m_u+m_s)m_d\right ]^{1/2}, \nonumber \\
 m_{K^0,{\bar K}^0} &=& \mp\frac{m_s^2-m_d^2}{2p_F}+\left [
       \frac{4A}{f_{\pi}^2}(m_s+m_d)m_u\right ]^{1/2}, \nonumber \\
\end{eqnarray}
with $A=3\Delta^2/(4\pi^2)$ \cite{CFL-Son,smass-Schafer}. It was
found that the kaon masses are substantially affected by the strange
quark mass, the masses of $K^-$ and ${\bar K}^0$ are pushed up while
$K^+$ and $K^0$ are lowered. As a result, the $K^+$ and $K^0$ become
massless if $m_s|_{crit} =3.03 ~ m_d^{1/3} \Delta^{2/3}$. For larger
values of $m_s$ the kaon modes are unstable, signaling the formation
of a kaon condensate. Recently, it was found that in the CFL phase,
there also may exist $\eta$ condensate \cite{CFLeta-Schafer}.

{\bf g2SC and gCFL:} When the $\beta$-equilibrium and the charge
neutrality condition are required for the two-flavor quark system,
the Fermi surfaces of the pairing $u$ quark and $d$ quark differ by
$\mu_e$, here $\mu_e$ is the chemical potential for electrons. It
was found that when the gap parameter $\Delta < \mu_e/2$, the system
will be in a new ground state called the gapless 2SC (g2SC) phase
\cite{g2SC-SH}. The g2SC phase has very unusual temperature
properties \cite{g2SC-HS-T} and chromomagnetic properties
\cite{chromo-ins-g2SC}. This phase will be introduced in more detail
in Sec.~\ref{g2sc}.

Similarly, for a charge neutral 3-flavor system with a nonzero strange quark mass $m_s$, with
increasing $m_s$, the CFL phase transfers to a new gapless CFL (gCFL) phase
when $m_s^2/\mu \simeq 2 \Delta$ \cite{gCFL-AKR}. The finite temperature property
of the charge neutral three-flavor quark matter was investigated in
Ref.~\cite{udSC-Hatsuda,gCFL-Ruster,gCFL-Kenji}. Recently, it was shown that the kaon
condensate shifts the critical strange quark mass to higher values for the
appearance of the gCFL phase \cite{gCFL-condensate}.

\subsection{Gapless color superconductor}
\label{g2sc}

In this section, I would like to focus on unconventional color
superconductor with mismatched pairing by taking charge neutral
two-flavor system as an example.

It is very likely that the color superconducting phase may exist in
the core of compact stars, where bulk matter should satisfy the
charge neutrality condition. This is because bulk matter inside the
neutron star is bound by the gravitation force, which is much weaker
than the electromagnetic and the strong color forces. Any electric
charges or color charges will forbid the formation of bulk matter.
In addition, matter inside neutron star also needs to satisfy the
$\beta$-equilibrium.

In the ideal two-flavor color superconducting (2SC) phase, the
pairing $u$ and $d$ quarks have the same Fermi momenta. Because $u$
quark carries electric charge $2/3$, and $d$ quark carries electric
charge $-1/3$, it is easy to check that quark matter in the ideal
2SC phase is positively charged. To satisfy the electric charge
neutrality condition, roughly speaking, twice as many $d$ quarks as
$u$ quarks are needed. This induces a large difference between the
Fermi surfaces of the two pairing quarks, i.e., $\mu_d - \mu_u =
\mu_e \approx \mu/4$, where $\mu,\mu_e$ are chemical potentials for
quarks and electrons, respectively. Naively, one would expect that
the requirement of the charge neutrality condition will destroy the
$ud$ Cooper pairing in the 2SC phase.

Indeed, the interest in the charge neutral 2SC phase was stirred by
the paper Ref.~\cite{absence2SC}. It was claimed in this paper that
there will be no 2SC phase inside neutron star under the requirement
of the charge neutrality condition. In fact, the authors meant that
for a charge neutral three flavor system, the 2SC+s phase is not
favorable compared to the CFL phase. This is a natural result under
the assumption of a {\it small} strange quark mass, even without the
requirement of the charge neutrality condition. In the framework of
the bag model, in which the strange quark mass is very small, the
CFL phase is always the ground state for cold dense quark matter,
and there is no space for the existence of two-flavor quark matter.

However, there is another scenario about the hadron-quark phase
transition in the framework of the SU(3) NJL model. In the vacuum,
quarks obtain their dynamical masses induced by the chiral
condensate. $u,d$ quarks have constituent mass around $330 {\rm
MeV}$, while the $s$ quark has heavier constituent mass, which is
around $500 {\rm MeV}$. With the increasing of the bayron density,
the constituent quark mass starts to decrease when the chemical
potential becomes larger than its vacuum constituent mass. In this
scenario, $s$ quark restores chiral symmetry at a larger critical
chemical potential than that of $u,d$ quarks. If the deconfinement
phase transition happens sequentially, there will exist some baryon
density regime for only $u,d$ quark matter and $s$ quark is still
too heavy to appear in the system.

It is worth to mention that the effect of the electric charge
neutrality condition on a three-flavor quark system is very
different from that on a two-flavor quark system. Because $s$ quark
carries $-1/3$ electric charges, it is much easier to neutralize the
electric charges in a three-flavor quark system than that in a
two-flavor quark system. However, the color charge neutrality
condition is nontrivial in a three-flavor quark system, when the
strange quark mass is not very small. For a  detailed consideration
of the charge neutral three-flavor system, see recent papers
Refs.~\cite{gCFL-AKR,udSC-Hatsuda,gCFL-Ruster,gCFL-Kenji}.

In the following, we focus on the charge neutral two flavor quark
system. Motivated by the sequential deconfinement scenario, the
authors of Ref.~\cite{N2SC-Steiner} investigated charge neutral
quark matter based on the SU(3) NJL model. To large extent, their
results agree with those in Ref.~\cite{absence2SC}, i.e., the CFL
phase is more favorable than the 2SC+s phase in charge neutral
three-flavor cold dense quark matter, and they did not find the
charge neutral 2SC phase.

However, it was found in Ref.~\cite{N2SC-HZC} that a charge neutral
two-flavor color superconducting (N2SC) phase does exist, which was
confirmed in Refs.~\cite{Diploma-Ruster,Mishra-Mishra}. Comparing
with the ideal 2SC phase, the N2SC phase found in
Ref.~\cite{N2SC-HZC} has a largely reduced diquark gap parameter,
and the pairing quarks have different number densities. The latter
contradicts the paring ansatz in Ref.~\cite{pairansatz}. Therefore,
one could suggest that this phase is an unstable Sarma state
\cite{Sarma}. In Ref.~\cite{g2SC-SH}, it was shown that the N2SC
phase is a stable state under the restriction of the charge
neutrality condition. As a by-product, which comes out as a very
important feature, it was found that the quasi-particle spectrum has
zero-energy excitation in this charge neutral two-flaovr color
superconducting phase. Thus this phase is named the ``gapless
2SC(g2SC)" phase.

%\subsubsection{Gapless color superconductor in the mean field approximation}

The 2-flavor system can be described by the gauged
Nambu--Jona-Lasinio (gNJL) model, the Lagrangian density has the
form of
\begin{eqnarray}
\label{lagr} {\cal L} &=& {\bar q}\Big ( i\fsl{D}+\hat{\mu}\gamma^0
\Big ) q  +
   G_S  \Big [(\bar qq)^2+
(\bar qi\gamma^5\vec\tau q)^2\Big ]  \nonumber \\
&+& G_D \Big [\bar q^Ci\gamma^5\tau_2 \epsilon^{\rho} q\Big ] \Big
[\bar qi\gamma^5\tau_2 \epsilon^{\rho} q^C\Big ] , \label{lg-2sc}
\end{eqnarray}
with $D_\mu \equiv \partial_\mu - ig A_\mu^{a} T^{a}$.  Here
$A_\mu^{a}$ are gluon fields,  $T^a=\lambda^a/2$ are the generators
of $SU(3)_{\rm c}$ gauge group with $a=1, \cdots, 8$.
%We have explicitly specified the anti-symmetric
% matrices $T^A$ with $A=2,5,7 \in a$.
In the gNJL model, the gauge fields are external fields and do not
contribute to the dynamics of the system. The property of the color
superconducting phase characterized by the diquark gap parameter is
determined by the nonperturbative gluon fields, which has been
simply replaced by the four-fermion interaction in the NJL model.
$G_S$ and $G_D$ are the quark-antiquark coupling constant and the
diquark coupling constant, respectively. $q^C=C {\bar q}^T$, ${\bar
q}^C=q^T C$ are charge-conjugate spinors, $C=i \gamma^2 \gamma^0$ is
the charge conjugation matrix (the superscript $T$ denotes the
transposition operation). The quark field $q \equiv q_{i\alpha}$
with $i=u,d$ and $\alpha=r,g,b$ is a flavor doublet and color
triplet, as well as a four-component Dirac spinor, ${\bf
\tau}=(\tau^1,\tau^2,\tau^3)$ are Pauli matrices in the flavor
space, where $\tau^2$ is antisymmetric, and $(\varepsilon)^{ik}
\equiv \varepsilon^{ik}$, $(\epsilon^b)^{\alpha \beta} \equiv
\epsilon^{\alpha \beta b}$ are totally antisymmetric tensors in the
flavor and color spaces.

In $\beta$-equilibrium, the matrix of chemical potentials in the
color-flavor space ${\hat \mu}$ is given in terms of the quark
chemical potential $\mu$, the chemical potential for the electrical
charge $\mu_e$ and the color chemical potential $\mu_8$,
\begin{eqnarray}
\mu_{ij}^{\alpha\beta} = (\mu \delta_{ij} - \mu_e
Q_{ij})\delta^{\alpha\beta}
 + \frac{2}{\sqrt{3}} \mu_8 \delta_{ij} (T_8)^{\alpha\beta}.
\end{eqnarray}

The total thermodynamic potential for $u, d$ quarks in
$\beta$-equilibrium with electrons takes the form
\cite{N2SC-HZC,g2SC-SH,g2SC-HS-T}:
\begin{eqnarray}
\Omega_{u,d,e} &=&
-\frac{1}{12\pi^2}\left(\mu_{e}^{4}+2\pi^{2}T^{2}\mu_{e}^{2}
+\frac{7\pi^{4}}{15} T^{4} \right) \nonumber \\
& + & \frac{m^2}{4G_S} +  \frac{\Delta^2}{4G_D} \nonumber\\
&-& \sum_{A} \int\frac{d^3 p}{(2\pi)^3} \left[E_{A} +2
T\ln\left(1+e^{-E_{A}/T}\right)\right] , \label{pot-ude}
\end{eqnarray}
where the electron mass was taken to be zero, which is sufficient
for the purposes of the current study. The sum over $A$ runs over
all (6 quark and 6 antiquark) quasi-particles. The explicit
dispersion relations and the degeneracy factors of the
quasi-particles read
\begin{eqnarray}
E_{ub}^{\pm} &=& E(p) \pm \mu_{ub} , \hspace{26.6mm} [\times 1]
\label{disp-ub} \\
E_{db}^{\pm} &=& E(p) \pm \mu_{db} , \hspace{26.8mm} [\times 1]
\label{disp-db}\\
E_{\Delta^{\pm}}^{\pm} &=& E_{\Delta}^{\pm}(p) \pm  \delta \mu .
\hspace{25.5mm} [\times 2] \label{2-degenerate}
\end{eqnarray}

\vskip 0.2cm {\bf Gapless excitation in quasi-particle spectrum}
\vskip 0.2cm

It is instructive to start with the excitation spectrum in the case
of the ideal 2SC phase when $\delta\mu=0$. With the conventional
choice of the gap pointing in the anti-blue direction in the color
space, the blue quarks are not affected by the pairing dynamics, and
the other four quarsi-particle excitations are linear superpositions
of $u_{r,g}$ and $d_{r,g}$ quarks and holes. The quasi-particle is
nearly identical with a quark at large momenta and with a hole at
small momenta. We represent the quasi-particle in the form of
$Q(quark, hole)$, then the four quasi-particles can be represented
explicitly as $Q(u_r, d_g)$, $Q(u_g, d_r)$, $Q(d_r, u_g)$ and
$Q(d_g, u_r)$. When $\delta\mu=0$, the four quasi-particles are
degenerate, and have a common gap $\Delta$. If there is a small
mismatch ($\delta\mu < \Delta$) between the Fermi surfaces of the
pairing $u$ and $d$ quarks, the excitation spectrum will change. It
is found that $\delta \mu$ induces two different dispersion
relations, the quasi-particle $Q(d_g, u_r)$ has a smaller energy gap
$\Delta - \delta\mu$, and the quasi-particle $Q(u_r, d_g)$ has a
larger energy gap $\Delta + \delta\mu$. This is similar to the case
when the mismatch is induced by the mass difference of the pairing
quarks \cite{gapless-ABR}.

\begin{figure}
\includegraphics[width=6cm]{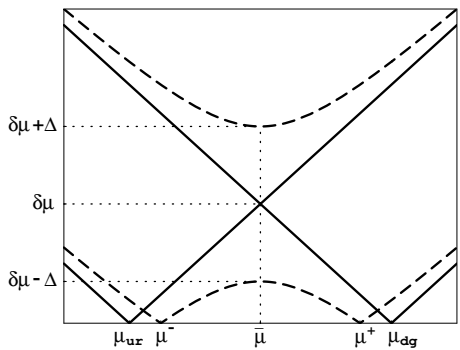}
\caption{Dispersion relation of the gapless phase. The figure is
taken from Ref. \cite{g2SC-SH}.} \label{fig-disp-gapless}
\end{figure}

If the mismatch $\delta\mu$ is larger than the gap parameter
$\Delta$, the lower dispersion relation for the quasi-particle
$Q(d_g, u_r)$ will cross the zero-energy axis, as shown in the right
panel of Fig.~\ref{fig-disp-gapless}. The energy of the
quasi-particle $Q(d_g, u_r)$ vanishes at two values of momenta
$p=\mu^{-}$ and $p=\mu^{+}$ where $\mu^{\pm}\equiv
\bar\mu\pm\sqrt{(\delta\mu)^2-\Delta^2}$. Thus this phase is called
the gapless 2SC (g2SC) phase.

%%%%%%%%%%%%%%%%%%%%%%%%%%%%%%%%%%%%
%%%%%%%%%%%%%%%%%%%%%%%%%%%%%%%%%%%%
%%%%%%%%%%%%%%%%%%%%%%%%%%%%%%%%%%%%

\vskip 0.2cm {\bf Thermal stable charge neutral g2SC state} \vskip
0.2cm

An unstable gapless CFL phase has been found in
Ref.~\cite{gapless-ABR}, and a similar stable gapless color
superconductivity could also appear in a cold atomic gas
\cite{Liu-Wilczek} or in $u, s$ or $d, s$ quark matter when the
number densities are kept fixed \cite{gapless-GLW}. Also, some
gapless phases may appear due to P-wave interactions in the cold
atomic system \cite{P-wave}. However, the gapless 2SC phase is a
thermal stable state under the charge neutrality condition.

If a macroscopic chunk of quark matter exists inside compact stars,
it must be neutral with respect to electric as well as color
charges. Now, we discuss the role of the electric charge neutrality
condition. If a macroscopic chunk of quark matter has nonzero net
electric charge density $n_Q$, the total thermodynamic potential for
the system should be given by
\begin{eqnarray}
\Omega &=&  \Omega_{Coulomb} + \Omega_{u,d,e},
\end{eqnarray}
where $\Omega_{Coulomb} \sim n_Q^2 V^{2/3}$ ($V$ is the volume of
the system) is induced by the repulsive Coulomb interaction. The
energy density grows with increasing the volume of the system, as a
result, it is almost impossible for matter inside stars to remain
charged over macroscopic distances. So bulk quark matter must
satisfy electric neutrality condition with
$\Omega_{Coulomb}|_{n_Q=0}=0$, and $\Omega_{u,d,e}|_{n_Q=0}$ is on
the neutrality line. Under the charge neutrality condition, the
total thermodynamic potential of the system is
$\Omega|_{n_Q=0}=\Omega_{u,d,e}|_{n_Q=0}$.

%%%%%%%%%%%%%%%%%%%%%%%%%%%
\begin{figure}
\vskip 0.3cm
\centerline{\epsfxsize=6cm\epsffile{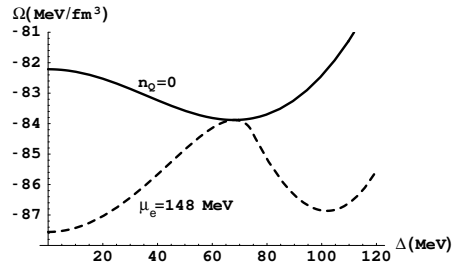}}
\caption{The effective potential as a function of the diquark gap
$\Delta$ calculated at a fixed value of the electric chemical
potential $\mu_e = 148 $ ~MeV (dashed line), and the effective
potential defined along the neutrality line (solid line). The
results are plotted for $\mu=400$ MeV with $\eta=0.75$. The figure
is taken from Ref. \cite{g2SC-SH}.} \label{V2D}
\end{figure}
%%%%%%%%%%%%%%%%%%%%%%%%%

Here, we want to emphasize that: The proper way to find the ground
state of homogeneous neutral $u, d$ quark matter is to minimize the
thermodynamic potential along the neutrality line $\Omega|_{n_Q=0} =
\Omega_{u,d,e}|_{n_Q=0}$. This is different from that in the flavor
asymmetric quark system, where $\beta$-equilibrium is required but
$\mu_e$ is a free parameter, and the ground state for flavor
asymmetric quark matter is determined by minimizing the
thermodynamic potential $\Omega_{u,d,e}$. At a fixed $\mu_e=148
~{\rm MeV}$ and with color charge neutrality, the thermodynamic
potential is shown as a function of the diquark gap by the dashed
line in Fig.~\ref{V2D}. The minimum gives the ground state of the
flavor asymmetric system, and the corresponding diquark gap is
$\Delta=0$, but this state has negative electric charge density, and
cannot exist in the interior of compact stars.

\subsection{Chromomagnetic instability in the g2SC phase}

As we know, one of the most important properties of the ordinary
superconductor is its Meissner effect, i.e., a superconductor expels
the magnetic field, which was discovered by Meissner and Ochsenfeld
in 1933 \cite{Meissner}. From theoretical point of view, the
Messiner effect can be explained using the linear response theory.
The induced current $j^{ind}_{i}$ is related to the magnetic field
$A_{j}$ by $j^{ind}_{i} = \Pi_{ij} A^{j} $, where the response
function $\Pi_{ij}$ is the photon polarization tensor. The response
function has two components, diamagnetic and paramagnetic part
\cite{PD-Nam}. In the static and long-wavelength limit, for the
normal metal, the paramagnetic component cancels exactly the
diamagnetic component. While in the superconducting phase, the
paramagnetic component is quenched by the energy gap and producing a
net diamagnetic response. Thus the ordinary superconductor is a
perfect diamagnet.

In color superconducting phases, the gluon self-energy (the response
function to an external color field), has been investigated in the
ideal 2SC phase \cite{Meissner2SC} and in the CFL phase
\cite{Meissner-CFL}. The results show that the gauge bosons
connected with the broken generators obtain masses in these phases,
which indicate the Meissner screening effect in these phases.

It is very interesting to know the chromomagnetic property in the
g2SC phase. We studied the g2SC phase in the framework of the SU(2)
NJL model, and the NJL model lacks gluons. As reflection of this, it
possesses the global instead of gauged color symmetry. In addition,
there appear five Nambu-Goldstone (NG) bosons in the ground state of
the model when the color symmetry is broken. In QCD, there is no
room for such NG bosons. However, the NJL model can be thought of as
the low energy theory of QCD in which the gluons, as independent
degrees of freedom, are integrated out. The gluons could be
reintroduced back by gauging the color symmetry in the Lagrangian
density of the NJL model, providing a semirigorous framework for
studying the effect of the Cooper pairing on the physical properties
of gluons.

The existence of the g2SC phase can be regarded as a physical and
model independent result under the restriction of local charge
neutrality condition, the order parameter for this phase is $\Delta
< \delta\mu$. In Ref.~\cite{chromo-ins-g2SC}, we calculated the
gluon self-energy in the g2SC phase. It is found that, in this
phase, the symmetry broken gauge bosons have imaginary Meissner
screening masses, which is induced by the dominant paramagnetic
contribution to the gluon self-energy. In condensed matter, this
phenomenon is called the paramagnetic Meissner effect(PME)
\cite{PME}, and has been observed in some high temperature
superconductors and small superconductors.

%%%%%%%%%%%%%%%%%%%%%%%%%%%%%%%%%%%%
\begin{figure}[t]
\includegraphics[width=7cm]{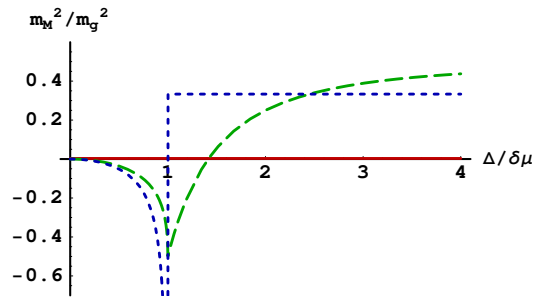}
\caption{ Squared values of the gluon Debye (upper panel) and
Meissner (lower panel) screening masses, devided by $m_g^2
=4\alpha_s\bar\mu^2/3\pi$, as functions of the dimensionless
parameter $\Delta/\delta\mu$. The red solid  line denotes the
results for the gluons with $A=1,2,3$, the green long-dashed line
denotes the results for the gluons with $A=4,5,6,7$, and the blue
short-dashed line denotes the results for the gluon with
$A=\tilde{8}$. The figure is taken from Ref.
\cite{chromo-ins-g2SC}.} \label{fig-Masses} \end{figure}
%%%%%%%%%%%%%%%%%%%%%%%%%%%%%%%%%%%%

Unavoidably, the imaginary Meissner screening mass indicates a
chromomagnetic instability of the g2SC phase.  There are many
proposals on resolving the chromomagnetic instability
\cite{Instability-NG,Instability-GluonCond,Instability-LOFF,Instability-Higgs,Instability-Amplitude}:
One is through a gluon condensate to stabilize the system, which may
not change the structure of the g2SC phase. It is also possible that
the instability drives a new stable ground state, which may have a
rotational symmetry breaking like in Refs.~\cite{DFS,DFS-1}, or even
have an inhomogeneous phase structure, like a crystal
\cite{LO,FF,LOFF-1} or a vortex \cite{votex} structure.

%%%%%%%%%%%%%%%%%%%%%%%%%
%%%%%%%%%%%%%%%%%%%%%%%%%

\subsection{Sarma instability and Higgs instability}

In order to understand the chromomagnetic instability in the gapless
phases, we extend the gauged NJL model beyond mean-field
approximation \cite{Instability-Higgs}. The superconducting state is
characterized by the order parameter $\Delta(x)$, which is a complex
scalar field and has the form of $\Delta (x) = |\Delta(x)|
e^{i\varphi (x)}$, with $|\Delta| $ the amplitude and $\varphi$ the
phase of the gap order parameter. For a homogeneous condensate,
$\Delta(x)$ is a spatial constant. The fluctuations of the phase
give rise to the pseudo Nambu-Goldstone boson(s), while that of the
amplitude to the Higgs field, following the terminology of the
electroweak theory. Stimulated by the role of the phase fluctuation
in the unconventional superconducting phase \cite{HTSC} in condensed
matter, we formulate the 2SC phase in the nonlinear realization
framework \cite{NonLinear} in order to naturally take into account
the contribution from the phase fluctuation or pseudo
Nambu-Goldstone current(s).

In the 2SC phase, the color symmetry $G=SU(3)_c$  breaks to
$H=SU(2)_c$. The generators of the residual $SU(2)_c$ symmetry H are
$\{S^a=T^a\}$ with $a=1,2,3$ and the broken generators
$\{X^b=T^{b+3}\}$ with $b=1, \cdots, 5$. (More precisely, the last
broken generator is a combination of $T_8$ and the generator ${\bf
1}$ of the global $U(1)$ symmetry of baryon number conservation, $B
\equiv ({\bf 1} + \sqrt{3} T_8)/3$ of generators of the global
$U(1)_B$ and local $SU(3)_c$ symmetry. )

The coset space $G/H$ is parameterized by the group elements
\begin{equation} \label{phase}
{\cal V}(x) \equiv \exp \left[ i \left( \sum_{a=4}^8 \varphi_a(x)
T_a \right) \right]  \,\,,
\end{equation}
here operator ${\cal V}$ is unitary, and ${\cal V}^{-1} = {\cal
V}^\dagger$ and $\varphi_a (a=4,\cdots,7)$ and $\varphi_8$ are five
Nambu-Goldstone diquarks, and we have not consider the topologically
nontrivial case and therefore ${\cal V}(x)$ can be expanded
uniformly according to the powers of $\varphi$'s. In fact, ${\cal
V}(x)$ is alway topologically trivial for a configuration of
$\varphi$'s that has a finite energy because of the trivial homotopy
group $\pi_2(SU(3)/SU(2))$.

Introducing a new quark field $\chi$, which is connected with the
original quark field $q$ in Eq. (\ref{lg-2sc}) through a nonlinear
transformation,
\begin{equation} \label{chi}
q = {\cal V}\, \chi \,\,\,\, , \,\,\,\,\, \bar{q} = \bar{\chi}\,
{\cal V}^\dagger\,\, ,
\end{equation}
and the charge-conjugate fields transform as
\begin{equation}
q_{C} = {\cal V}^* \, \chi_{C} \,\,\,\, , \,\,\,\,\, \bar{q}_{C} =
\bar{\chi}_{C} \, {\cal V}^T\,\, .
\end{equation}
The advantage of transforming the quark fields is that this
preserves the simple structure of the terms coupling the quark
fields to the diquark sources,
\begin{equation}
\bar{q}_{C}\, \Delta^+ \, q \equiv \bar{\chi}_{C}\, \Phi^+ \, \chi
\,\,\,\, , \,\,\,\,\, \bar{q}\, \Delta^- \, q_{C} \equiv \bar{\chi}
\, \Phi^- \, \chi_{C} \,\, .
\end{equation}

In the Nambu-Gor'kov space of the new spinors
\begin{equation}
X \equiv \left( \begin{array}{c}
                    \chi \\
                    \chi_{C}
                   \end{array}
            \right) \,\,\, , \,\,\,\,
\bar{X} \equiv ( \bar{\chi} \, , \, \bar{\chi}_{C} ),
\end{equation}
the nonlinear realization of the original Lagrangian density
Eq.~(\ref{lg-2sc}) takes the form of
\begin{equation}
{\cal L}^{nl}_{2SC} \equiv - \frac{\Phi^+\Phi^-}{4 G_D} \,\, +
\frac{1}{2} \bar{X} \,  {\cal S}_{nl}^{-1} \, X , \label{lagr-nl}
\end{equation}
with
\begin{equation}
{\cal S}_{nl}^{-1} \equiv \left( \begin{array}{cc}
            [G^+_{0,nl}]^{-1} & \Phi^- \\
             \Phi^+ & [G^-_{0,nl}]^{-1}
       \end{array} \right)\,\, .
\end{equation}
Here the explicit form of the free propagator for the new quark
field is
\begin{eqnarray}
[G^+_{0,nl}]^{-1} & = & i\, \fsl{D} + {\hat \mu} \, \gamma_0 +
\gamma_\mu \, V^\mu,
%[G^+_{0,nl}]^{-1} & = & i\, \gamma^\mu \partial_\mu + {\hat \mu} \, \gamma_0 + \gamma_\mu \, V^\mu,
\end{eqnarray}
and
\begin{eqnarray}
[G^-_{0,nl}]^{-1} & = & i\, \fsl{D}^T - {\hat \mu} \, \gamma_0 +
\gamma_{\mu} \, V_C^\mu .
\end{eqnarray}
Comparing with the free propagator in the original Lagrangian
density, the free propagator in the non-linear realization framework
naturally takes into account the contribution from the
Nambu-Goldstone currents or phase fluctuations, i.e.,
\begin{eqnarray}
V^\mu & \equiv & {\cal V}^\dagger \, \left( i \, \partial^\mu \right) \, {\cal V}, \nonumber \\
V^\mu_C & \equiv & {\cal V}^T \, \left( i \, \partial^\mu \right) \,
{\cal V}^*,
\end{eqnarray}
which is the $N_c N_f \times N_c N_f$-dimensional Maurer-Cartan
one-form introduced in Ref. \cite{NonLinear}. The linear order of
the Nambu-Goldstone currents $V^\mu$ and $V_C^\mu $ has the explicit
form of
\begin{eqnarray}
V^\mu & \simeq &  - \sum_{a=4}^8
\left( \partial^\mu  \varphi_a \right)\, T_a  \,\, , \\
V_C^\mu & \simeq &   \sum_{a=4}^8
 \left(\partial^\mu \varphi_a \right) \, (T_a)^T  \,\, .
\end{eqnarray}

The advantage of the non-linear realization framework Eq.
(\ref{lagr-nl}) is that it can naturally take into account the
contribution from the phase fluctuations or Nambu-Goldstone
currents. The task left is to find the correct ground state by
exploring the stability against the fluctuations of the magnitude
and the phases of the order parameters. The free energy
$\Omega(V_{\mu},\Phi, \mu,\mu_8, \mu_e)$ can be evaluated
 directly and it  takes the form of

\begin{eqnarray}
\Omega_{nl}(V_{\mu}, \Phi, \mu, \mu_8, \mu_e) &=& - \frac{1}{2}
T\sum_n\int\frac{d^3\vec{p}}{(2\pi)^3} {\rm Tr} \ln ( [{\cal
S}_{nl}(P)]^{-1}) \nonumber \\
& + & \frac{\Phi^2}{4 G_D}.
%\label{free-energy-NG}
\end{eqnarray}

In the 2SC phase, the color symmetry $SU(3)_c$  is spontaneously
broken to $SU(2)_c$ and diquark field obtains a nonzero expectation
value. Without loss of generality, one can always assume that
diquark condenses in the anti-blue direction, i.e., only red and
green quarks participate the Cooper pairing, while blue quarks
remains as free particles. The ground state of the 2SC phase is
characterized by  $\vev{\Delta^{3}}\equiv \Delta$, and
$\vev{\Delta^{1}}=0$, $\vev{\Delta^{2}}=0$.

Considering the fluctuation of the order parameter,  the diquark
condensate can be parameterized as
\begin{eqnarray}
\vect\Delta^1(x)\\ \Delta^2(x)\\ \Delta^3(x)\evect &=& \exp \left[ i
\left( \sum_{a=4}^8 \varphi_a(x) T_a
 \right) \right] \vect0\\0\\ \Delta + H(x) \evect \nonumber \\
&\equiv &{\cal V}(x) \Phi^{\rho}(x),
% = \frac 1{\sqrt{2}}\vect \xi_5-i\xi_4 \\ \xi_7-i\xi_6 \\ \sqrt{2}\Delta
% + H +i\frac 2{\sqrt{3}}\xi_8\evect +o(\xi_i,\eta),
\label{II}
\end{eqnarray}
where $\Phi^{\rho}(x)= (0, \, 0, \, \Delta + H(x))$ is the diquark
field in the nonlinear realization framework, $\varphi_a, \varphi_8$
are Nambu-Goldstone bosons, and $H$ is the Higgs field.

Expanding the diquark field $\Phi^{\rho}$ around the ground state:
$\Phi^{\rho}= (0, \, 0, \, \Delta)$, the free-energy of the system
takes the following expression as
\begin{eqnarray}
\Omega_{nl} = \Omega_M +\Omega_{NG} + \Omega_{H}.
\label{free-energy-NG}
\end{eqnarray}
There are three contributions to the free-energy, the mean-field
approximation free-energy part $\Omega_M$ has the form of
\begin{equation}
\Omega_M= - \frac{T}{2} \sum_n\int\frac{d^3\vec{p}}{(2\pi)^3} {\rm
Tr} \ln ( [{\cal S}_M(P)]^{-1}) + \frac{\Delta^2}{4 G_D},
\end{equation}
the free-energy from the Higgs field $\Omega_H$ has the form of
\begin{eqnarray}
\Omega_H = \frac{T}{2}\sum_{k_0}\int\frac{d^3\vec
k}{(2\pi)^3}H^*(K)\Pi_{H}(K)H(K) \label{higgs}
\end{eqnarray}
with
\begin{eqnarray}
\Pi_{H}(K)&=&\frac{1}{2G_D} -
\frac{T}{2}\sum_{p_0}\int\frac{d^3\vec{p}}{(2\pi)^3} {\rm Tr}\Big [
{\cal S}_M(P+K)\, \nonumber \\
& & \left(\begin{array}{cc}
0 & i \tau_2\epsilon^{3} \gamma_5 \\
-i \tau_2\epsilon^{3} \gamma_5 & 0
\end{array}\right) \, \nonumber \\
& & {\cal S}_M(P) \, \left(\begin{array}{cc}
0 & i \tau_2\epsilon^{3} \gamma_5 \\
-i \tau_2\epsilon^{3} \gamma_5 & 0
\end{array}\right)\Big],
\end{eqnarray}
and the free-energy from the Nambu-Goldstone currents $\Omega_{NG}$
has the form of
\begin{eqnarray}
\Omega_{NG} & = &  - \frac{T^2}{4} \sum_{k_0}
\sum_{p_0}\int\frac{d^3\vec{k}}{(2\pi)^3}
\frac{d^3\vec{p}}{(2\pi)^3} \nonumber \\
& & {\rm Tr} \Big [ {\cal S}_M(P+K)\, \left(\begin{array}{cc}
\omega^\mu(-K)\gamma_\mu & 0 \\
0 & \omega_C^\mu(-K)\gamma_\mu
\end{array}\right) \, \nonumber \\
& & {\cal S}_M(P) \, \left(\begin{array}{cc}
\omega^\mu(K)\gamma_\mu & 0 \\
0 & \omega_C^\mu(K)\gamma_\mu
\end{array}\right) \, \Big ],
\label{goldstone}
\end{eqnarray}
with
\begin{eqnarray}
\omega^\mu(K) & = & g \, A^\mu_a(K) \, T_a - V^{\mu}(K) \,\, , \\
\omega_C^\mu(K) & =& - g \, A^\mu_a(K) \, T_a^T + V_C^{\mu}(K)\,\, .
\end{eqnarray}
where the inverse propagator ${\cal S}_M^{-1}$ takes the form of
\begin{eqnarray}
\left[{\cal S}_M(P)\right]^{-1} & = & \left(\begin{array}{cc}
\left[G_0^{+}(P)\right]^{-1} & i \tau_2\epsilon^{3} \gamma_5 \Delta \\
-i \tau_2\epsilon^{3} \gamma_5 \Delta & \left[G_0^{-}(P)\right]^{-1}
\end{array}\right) \nonumber \\
& = & \gamma^0(p_0+\rho_3\hat\mu)-\vec\gamma\cdot\vec
p+\Delta\rho_2\tau_2\epsilon^3\gamma_5. \nonumber \\
\label{prop-0}
\end{eqnarray}
The quasi-quark propagator at mean-field approximation has the form
of
\begin{equation}
{\cal S}_M = \left(\begin{array}{cc}
G^{+} & \Xi^{-} \\
\Xi^{+} & G^{-}
\end{array}\right).
\label{quarkpropagator}
\end{equation}
and its explicit expression of the Nambu-Gorkov components of ${\cal
S}_M$ has been derived in Ref. ~\cite{chromo-ins-g2SC}.

The Matsubara self-energy functions $\Pi_H(K)$ of the Higgs field
and that of the Goldstone fields (obtained after the sum over $p_0$
and integral over $\vec p$ in Eq.~(\ref{goldstone})) can be
continuated to real frequency following the standard procedure.
Because of the gapless excitations, the values of these functions at
zero frequency and zero mometum depends the order of the limit. In
this work, we shall restrict our attention to static fluctuations
only which amounts to replace $H(K)$, $\vec A(K)$ and $\varphi(K)$
of Eqs.~(\ref{higgs}) and (\ref{goldstone}) by $\sqrt{T}H(\vec
k)\delta_{k_0,0}$, $\sqrt{T}\vec A(\vec k)\delta_{k_0,0}$ and
$\sqrt{T}\varphi(\vec k)\delta_{k_0,0}$. Thus the long wavelength
limit of the self-energy functions discussed below corresponds to
the limit $\lim_{\vec k\to 0}\lim_{k_0\to 0}$.

\vskip 0.5cm {\bf Sarma instability} \vskip 0.5cm

In the mean-field approximation, the free-energy for $u, d$ quarks
in $\beta$-equilibrium takes the form \cite{g2SC-SH}:
\begin{eqnarray}
\Omega_M =
 \frac{\Delta^2}{4G_D}
-\sum_{A} \int\frac{d^3 p}{(2\pi)^3} \left[E_{A} +2
T\ln\left(1+e^{-E_{A}/T}\right)\right] . \nonumber \\
\label{pot-m}
\end{eqnarray}
which is the same as Eq.(\ref{pot-ude}) by neglecting the
contribution of free electrons and taking zero quark mass in the
color superconducting phase.

At zero temperature, the mean-field free-energy has the expression
of
\begin{eqnarray}
\Omega_M &=& \frac{\Delta^2}{4G_D} -\frac{\Lambda^4}{2\pi^2}
-\frac{\mu_{ub}^4}{12 \pi^2}
-\frac{\mu_{db}^4}{12 \pi^2} \nonumber \\
&-& 2 \int_{0}^{\Lambda} \frac{p^2 d p}{\pi^2} \left(\sqrt{(p+
\bar{\mu})^2+\Delta^2}
+\sqrt{(p-\bar{\mu})^2+\Delta^2}\right)\nonumber \\
&-& 2\theta \left(\delta\mu-\Delta\right)
\int_{\mu^{-}}^{\mu^{+}}\frac{p^2 d p}{\pi^2}\Big(
\delta\mu-\sqrt{(p-\bar{\mu})^2+\Delta^2} \Big). \nonumber
\\
\label{pot-2sc}
\end{eqnarray}

As we already knew that, with the increase of mismatch, the ground
state will be in the gapless 2SC phase when $\Delta < \delta\mu$,
the thermodynamical potential of which is given by
\begin{equation}
\Omega_M\simeq \Omega_M^{(0)}+\frac{2\bar\mu^2}{\pi^2}
\Big(\ln\frac{\delta\mu+\sqrt{\delta\mu^2-\Delta^2}}{\Delta_0}
-\delta\mu\sqrt{\delta\mu^2-\Delta^2}+\delta\mu^2\Big)
\label{omegaM}
\end{equation}
where $\Omega_M^{(0)}$ is the normal phase thermodynamic potential.
$\Delta_0$ the solution to the gap equation in the absence of
mismatch, $\delta\mu=0$. The solution to the gap equation reads
\begin{equation}
\Delta=\sqrt{\Delta_0(2\delta\mu-\Delta_0)}. \label{gapsol}
\end{equation}

The gapless phase is in principle a metastable Sarma state
\cite{Sarma}, i.e., the free-energy is a local maximum with respect
to the gap parameter $\Delta$. We have
\begin{equation}
\Big(\frac{\partial^2\Omega_M}{\partial\Delta^2}\Big)_{\bar\mu,\delta\mu}
=\frac{4\bar\mu^2}{\pi^2}\Big(1-\frac{\delta\mu}{\sqrt{\delta\mu^2-\Delta^2}}\Big).
\label{sarmains}
\end{equation}
The weak coupling approximation is employed in deriving
Eqs.~(\ref{omegaM}) and (\ref{sarmains}) from Eq.~(\ref{pot-2sc}),
which assumes that $\Delta_0$, $\Delta$ and $\delta\mu$ are much
smaller than $\mu$ and $\Lambda-\mu$. The same approximation will be
applied throughoutthe paper.

\vskip 0.5cm {\bf Nambu-Goldstone currents generation and the LOFF
state} \vskip 0.5cm

The quadratic action of the Goldstone modes in the long wavelength
limit can be written down with the aid of the Meissner masses
evaluated in Ref.\cite{chromo-ins-g2SC}. We find that
\begin{eqnarray}
\label{NG-free-energy} \Omega_{NG}&=&\frac{1}{2}\int d^3\vec
r\sum_{a=1}^8 m_a^2 ({\vec {\bf A}}^a -\frac{1}{g}\, {\vec
\triangledown}  \varphi^a)( {\vec {\bf A}}^a- \frac{1}{g}\, {\vec
\triangledown}  \varphi^a)\nonumber \\
 &+& higher \, orders \, .
\end{eqnarray}
where $m_1=m_2=m_3=0$,
\begin{eqnarray}
m_4^2&=&m_5^2=m_6^2=m_7^2=\frac{g^2\bar\mu^2}{3\pi^2}
\Big[\frac{\Delta^2-2\delta\mu^2}{2\Delta^2} \nonumber \\
&+&\theta(\delta\mu-\Delta)\frac{\delta\mu\sqrt{\delta\mu^2-\Delta^2}}{\Delta^2}\Big]
\end{eqnarray}
and
\begin{equation}
m_8^2=\frac{g^2\bar\mu^2}{9\pi^2}\Big[1-\frac{\delta\mu\theta(\delta\mu-\Delta)}{\sqrt{\delta\mu^2-\Delta^2}}\Big].
\end{equation}
It was found that at zero temperature, with the increase of
mismatch,  for five gluons with $a=4,5,6,7,8$ corresponding to
broken generator of $SU(3)_c$, their Meissner screening mass squares
become negative \cite{chromo-ins-g2SC}. This indicates the
development of the condensation of
\begin{equation}
\sum_{a=4}^{8} <{\vec {\bf A}}^a- \frac{1}{g}\, {\vec \triangledown}
\varphi^a> \neq 0.
% {\vec {\bf A}}^a - \frac{i}{g}\, {\vec \triangledown}  \varphi^a  \neq 0
\end{equation}
It can be interpreted as the spontaneous generation of
Nambu-Goldstone currents $\sum_{a}^{ 8}  <{\vec \triangledown}
\varphi^a> \neq 0$ \cite{Instability-NG}, or gluon condensation
$\sum_{a=4}^{8} <{\vec {\bf A}}^a> \neq 0$
\cite{Instability-GluonCond}. It can also be interpreted as a
colored-LOFF state \cite{Instability-LOFF} with the plane-wave order
parameter
\begin{equation}
\Delta(x)= \Delta {\rm e}^{ i \sum_{a=4}^{8}  {\vec \triangledown}
\varphi^a \cdot {\vec {\bf x}}}.
\end{equation}

\vskip 0.5cm {\bf Higgs instability} \vskip 0.5cm

We discussed the two known instabilities induced by mismatch, i.e.,
the Sarma instability and chromomagnetic instability, respectively.
It is found that there is another instability, which is related to
the Higgs field, and we call this instability "Higgs instability".

The free-energy from the Higgs field can be evaluated and takes the
form of
\begin{eqnarray}
\label{free-energy-H} \Omega_{H} & =&
\frac{T}{2}\sum_{k_0}\int\frac{d^3\vec{k}}{(2\pi)^3}
 H^*(\vec k)\Pi_{H}(k) H(\vec k).
% &=& \frac{1}{2} [(A_H+B_Hk^2)H^*H.
\end{eqnarray}

\begin{figure}
\includegraphics[width=7.5cm]{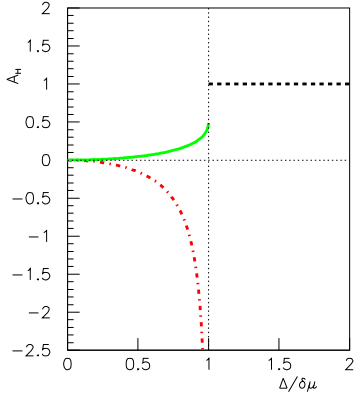}
\caption{ Squared values of the gluon Debye (upper panel) and
Meissner (lower panel) screening masses, devided by $m_g^2
=4\alpha_s\bar\mu^2/3\pi$, as functions of the dimensionless
parameter $\Delta/\delta\mu$. The red solid  line denotes the
results for the gluons with $A=1,2,3$, the green long-dashed line
denotes the results for the gluons with $A=4,5,6,7$, and the blue
short-dashed line denotes the results for the gluon with
$A=\tilde{8}$. This figure is taken from
Ref.\cite{Instability-Higgs}.} \label{fig-Higgs}
\end{figure}

\begin{figure}
\includegraphics[width=10cm]{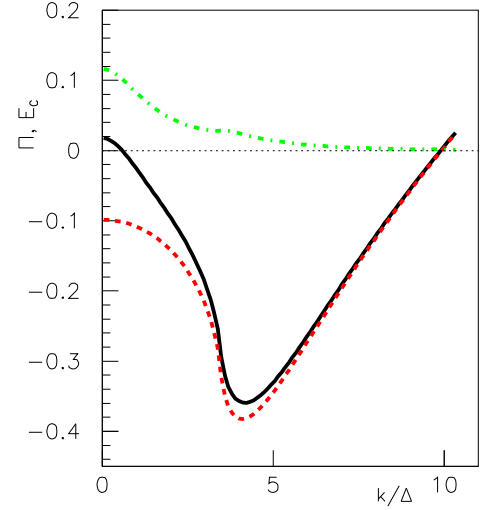}
\caption{\label{Higgs-k-fig} The function $\Pi_H(k)$ (red dashed
line), $\tilde\Pi_H(k)$ (black solid line) and the Coulomb energy
(green dash-dotted line) as functions of scaled-momentum $k/\Delta$,
in the case of $\delta\mu=2\Delta$ and
$(e^2\bar\mu^2)/(4\pi\Delta^2)=1$. This figure is taken from
Ref.\cite{Instability-Higgs}.} \label{fig-Higgs-k}
\end{figure}

Evaluating the one-loop quark-quark bubble $\Pi_H(k)$, we obtain
that
\begin{equation}
\Pi_H(k)=\frac{2\bar\mu^2}{\pi}I(k|\delta\mu)+\frac{\bar\mu^2}{2\pi\Delta^2}k^2J(k|\delta\mu),
\label{higgsenergy}
\end{equation}
where the functions $I(k|\delta\mu)$ and $J(k|\delta\mu)$ are given
by
\begin{eqnarray}
I(k|\delta\mu) &=& \Delta^2T
\sum_{n}\frac{1}{\sqrt{(\omega_n+i\delta\mu)^2+\Delta^2}} \nonumber
\\
& &
\int_{-1}^1dx\frac{1}{(\omega_n+i\delta\mu)^2+\Delta^2+\frac{1}{4}k^2x^2},
\label{Iintegral}
\end{eqnarray}
\begin{eqnarray}
J(k|\delta\mu)& = &\Delta^2T
\sum_{n}\frac{1}{\sqrt{(\omega_n+i\delta\mu)^2+\Delta^2}} \nonumber
\\
& &
\int_{-1}^1dx\frac{x^2}{(\omega_n+i\delta\mu)^2+\Delta^2+\frac{1}{4}k^2x^2}.
\label{Jintegral}
\end{eqnarray}
with ${\rm Re}\sqrt{(\omega_n+i\delta\mu)^2+\Delta^2}>0$. Here, the
summation over $n$ is the frequency summation at finite temperature
field theory, with $\omega_n=(2n+1)\pi T$. At T=0, the free energy
of the Higgs field has the form of
\begin{equation}
\Pi_H(k)=A_H+B_Hk^2
\end{equation}
with
\begin{equation}
A_H =
\left(\frac{\partial^2\Omega_M}{\partial\Delta^2}\right)_{\delta\mu}
=\frac{4\bar\mu^2}{\pi^2}\Big(1-\frac{\delta\mu}{\sqrt{(\delta\mu)^2-\Delta^2}}\Big),
\label{AH}
\end{equation}
\begin{equation}
B_H=\frac{2\bar\mu^2}{9\pi^2\Delta^2}\Big[1-\frac{(\delta\mu)^3}{((\delta\mu)^2-\Delta^2)^{\frac{3}{2}}}
\Big]. \label{BH}
\end{equation}
It follows from Eqs. (\ref{AH}) and  (\ref{BH}) that the Higgs field
becomes unstable in the gapless phase when $\delta\mu > \Delta$,
$A_H$ in the gapless phase is shown in Fig. \ref{fig-Higgs} by the
red dash-dotted line. The Higgs instability was also considered in
Ref. \cite{Instability-Amplitude}, where it was called as "amplitude
instability". It has to be pointed out that we got different
expressions for the coefficients of the gradient term. For
$k>>\Delta$, we have
\begin{equation}
\Pi_H(k)\simeq \frac{\bar\mu^2}{2\pi^2}\Big(2\ln\frac{k}{\Delta}-2
-\ln\frac{\delta\mu-\sqrt{\delta\mu^2-\Delta^2}}{\delta\mu+\sqrt{\delta\mu^2-\Delta^2}}\Big).
\end{equation}
Therefore the Higgs instability disappears for sufficiently large
momentum. The form factor $\Pi_H(k)$ for arbitrary momentum is
plotted using red dashed line in Fig. \ref{Higgs-k-fig} for a
typical value of the mismatch parameter $\Delta/\delta\mu=1/2$. We
notice that the Higgs instability becomes stronger for nonzero
momentum.

%%%%%%%%%%%%%%%%%%%%%%%%%%%%%%%%%%%%
%%%%%%%%%%%%%%%%%%%%%%%%%%%%%%%%%%%%

\vskip 0.5cm {\bf Charge neutrality condition} \vskip 0.5cm

Unless a competing mechanism that results in a positive contribution
to the inhomogeneous blocks of the stability matrix. The Higgs
instability will prevent gapless superfluidity/ superconductivity
from being implemented in nature. In the system of imbalanced
neutral atoms, such a mechanism is not likely to exist and this
contributes to the reason why the BP state has never been observed
there. For the quark matter being considered, however, the positive
Coulomb energy induced by the Higgs field of electrically charged
diquark pairs has to be examined.

The second order derivative of the Hemlholtz free energy for g2SC
reads:
\begin{equation}
\Big(\frac{\partial^2{\cal F}}{\partial\Delta^2}\Big)_{n_e} =
\Big(\frac{\partial^2\Omega}{\partial\Delta^2}\Big)_{\mu_e}
+\frac{\Big(\frac{\partial n_e}{\partial\Delta}\Big)_{\mu_e}^2}
{\Big(\frac{\partial n_e}{\partial\mu_e}\Big)_{\Delta}} \label{free}
\end{equation}
and the stability implemented by the charge neutrality implies that
\begin{equation}
\Big(\frac{\partial^2{\cal F}}{\partial\Delta^2}\Big)_{n_e}>0.
\end{equation}

While the global charge neutrality is maintained, an inhomogeneous
$\phi$ will induce local charge distribution. The corresponding
Coulomb energy
\begin{equation}
E_{\rm coul.}=\frac{1}{2}\sum_{\vec k\neq 0}\frac{\delta\rho(\vec
k)^*\delta\rho(\vec k)} {k^2+m_D^2(k)}, \label{coulomb}
\end{equation}
should be considered where $\delta\rho(\vec k)$ is the Fourier
component of the $\phi$-induced charge density and $m_D$ is the
Coulomb polarization function ( Debye mass at $\vec k=0$ ). We have
\begin{equation}
m_D^2(k)=-\frac{e^2T}{2}\sum_P{\rm tr}\gamma_0Q{\cal
S}(P+K)\gamma_0Q {\cal S}(P)
\end{equation}
and
\begin{equation}
\delta\rho(\vec k)=\kappa(k)H(\vec k),
\end{equation}
where
\begin{equation}
\kappa(k)=\frac{ieT}{2}\sum_P{\rm tr}\gamma_0Q{\cal S}(P+K)
\gamma_5\epsilon^3\rho_2{\cal S}(K)
\end{equation}
with $K=(0,\vec k)$ and $Q$ the electric charge operator. We have
\begin{equation}
Q=\rho_3(a+b\tau_3)
\end{equation}
with $a=1/6$ and $b=1/2$ for the quark matter consisting of $u$ and
$d$ flavors.

The modified Higgs self-energy has the form of
\begin{eqnarray}
\tilde\Pi_H(k) &\equiv& \Big(\frac{\partial^2{\cal F}}{\partial
H^*(\vec k)\partial H(\vec k)}\Big)_{n_e} \nonumber \\
& = &\Big(\frac{\partial^2\Omega}{\partial H^*(\vec k)\partial
H(\vec k)}\Big)_{\mu_e} +\frac{\kappa^*(k)\kappa(k)}{k^2+m_D^2(k)}
\nonumber \\
&=&\Pi_{H}(k)+\frac{\kappa^*(k)\kappa(k)}{k^2+m_D^2(k)}.
\label{higgscoul}
\end{eqnarray}
The stability of the system with respect to the Higgs field requires
that $\tilde\Pi_H(k)>0$ for all $k$.

The form factor $\kappa(q)$ and the momentum dependent Debye mass
square can be calculated explicitly at weak coupling, i.e.
$\delta\mu<<\bar\mu$ and $k<<\bar\mu$. We find that
\begin{equation}
\kappa(k)=\frac{2e^2\bar\mu^2b}{\pi}K(k|\delta\mu)
\end{equation}
with
\begin{eqnarray}
K(k|\delta\mu) &=& i\Delta T\sum_n\frac{\omega_n+i\delta\mu}
{\sqrt{(\omega_n+i\delta\mu)^2+\Delta^2}}\nonumber \\
& & \int_{-1}^1dx
\frac{1}{(\omega_n+i\delta\mu)^2+\Delta^2+\frac{1}{4}k^2x^2}
\label{Kintegral}
\end{eqnarray}
and
\begin{equation}
m_D^2(k)=\frac{6(a^2+b^2)e^2\bar\mu^2}{\pi^2}-\frac{2b^2e^2\bar\mu^2}{\pi}I(k|\delta\mu)
\end{equation}
with the first term the Debye mass of the normal phase and
$I(q|\delta\mu)$ the function defined in the section III.

\vskip 4mm {\bf Sarma instability at zero momentum limit can be
removed by Coulomb energy:} \vskip 4mm

Notice that
\begin{equation}
\lim_{k\to 0}\lim_{V\to\infty}\Big(\frac{\partial^2\Omega}{\partial
H^*(\vec k)\partial H(\vec k)}\Big)_{\mu_Q} =
\Big(\frac{\partial^2\Omega}{\partial\Delta^2}\Big)_{\mu_Q},
\end{equation}
\begin{equation}
m_D^2(0)=e^2\Big(\frac{\partial
n_e}{\partial\mu_e}\Big)_{\Delta,\mu_B}
\end{equation}
and
\begin{equation}
\kappa(0)=e\Big(\frac{\partial
n_e}{\partial\Delta}\Big)_{\mu_e,\mu_B}.
\end{equation}
We have
\begin{equation}
\lim_{\vec k\to 0}\Big(\frac{\partial^2{\cal F}}{\partial H^*(\vec
k)\partial H(\vec k)}\Big)_{\mu_e} = \Big(\frac{\partial^2{\cal
F}}{\partial\Delta^2}\Big)_{\mu_e},
\end{equation}
and the charge neutrality stabilize also the inhomogeneous Higgs
field with the momentum much smaller than the inverse coherence
length and the inverse Debye length.

In the static long-wave length limit, we have
\begin{equation}
m_D^2(0)=\frac{2e^2b^2\bar\mu^2}{\pi^2}\Big(1+\frac{2\delta\mu}
{\sqrt{\delta\mu^2-\Delta^2}}\Big) \label{debye0}
\end{equation}
and
\begin{equation}
\kappa(0)=\frac{4eb\bar\mu^2}{\pi^2}\frac{\Delta}{\sqrt{\delta\mu^2-\Delta^2}}.
\label{susc0}
\end{equation}
It follows from Eqs. (\ref{sarmains}), (\ref{higgscoul}),
(\ref{debye0}) and (\ref{susc0}) that the Higgs self-energy
including Coulomb energy correction
\begin{eqnarray}
& & {\tilde A}_H  \equiv  \tilde\Pi_{H}(0) =
\Big(\frac{\partial^2{\cal F}}{\partial\Delta^2}\Big)_{\mu,n_Q}
\nonumber
\\
&=& \frac{4(b^2-3a^2)\bar\mu^2
(\delta\mu-\sqrt{\delta\mu^2-\Delta^2})}{\pi^2[3a^2\sqrt{\delta\mu^2-\Delta^2}
+b^2(2\delta\mu+\sqrt{\delta\mu^2-\Delta^2})]}.
\nonumber \\
\end{eqnarray}
is always positive for the whole range of g2SC state. It means that
the Sarma instability in the gapless phase can be cured by Coulomb
energy. This is shown in Fig.~\ref{fig-Higgs}, where the red
dash-dotted line indicates the $A_H$ and the green solid line
indicates ${\tilde A}_H$.

\vskip 4mm {\bf  Higgs instability at nonzero momentum cannot be
removed by Coulomb energy: } \vskip 4mm

While the Sarma instability in g2SC phase can be cured by Coulomb
energy under the constraint of charge neutrality condition, it is
{\it{not sufficient}} for the system to be stable, even if the
chromomagnetic instabilities are removed, say by gluon condensation.
One has to explore the Higgs instability by calculating the
self-energy function $\tilde\Pi(k)$ in the whole momentum space,
which amounts to value the three basic functions $I(k|\delta\mu)$,
$J(k|\delta\mu)$ and $K(k|\delta\mu)$ defined in (\ref{Iintegral}),
(\ref{Jintegral}) and (\ref{Kintegral}).

Fig.~\ref{Higgs-k-fig} shows the Higgs self-energy $\Pi_H(k)$ (the
red dashed line) the Coulomb corrected Higgs self-energy ${\tilde
\Pi}_H(k)$ (the black solid line) and the Coulomb energy $E_{coul}$
(the green dash-dotted line) as functions of scaled-momentum
$k/\Delta$,  in the case of $\delta\mu=2\Delta$ and
$(e^2\bar\mu^2)/(4\pi\Delta^2)=1$.

Eventhough the Higgs instability can be removed by the Coulomb
energy for small momenta, it returns for intermediate momenta. This
phenomenon persists for a wide range of gap magnitude,
$0<\Delta<0.866\delta\mu$ and for all strength of the Coulomb
interaction, measured by the dimensionless ration
$\eta\equiv\frac{\alpha_e\bar\mu^2}{\Delta^2}$. Within the narrow
range $0.866\delta\mu<\Delta<\delta\mu$, the Higgs instability could
be removed if the Coulomb interaction were sufficiently strong. For
$\eta=1$, We found that $\tilde\Pi_H(k)>0$ for all $k$ if
$0.998\delta\mu<\Delta<\delta\mu$. In terms of the values of the
parameters of NJL model, we have $\alpha_e\bar\mu^2<\delta\mu$.
Therefore the electric Coulomb energy cannot cure the Higgs
instability for a realistic two flavor quark matter.

Negative $\Pi_H(k)$ indicates the Higgs mode is unstable and will
decay \cite{Coleman-Weinberg}. It is noticed that $\Pi_H(k)$ reaches
its minimum at a momentum, i.e., $k \simeq 4 \Delta$, which
indicates that a stable state may develop around this minimum, we
characterize this momentum as $k_{min}$. The inverse $k_{min}^{-1}$
is the typical wavelength for the unstable mode \cite{Weinberg-Wu}.
If mixed phase can be formed, the typical size $l$ of the 2SC
bubbles should be as great as $k_{min}^{-1}$, i.e., $l \simeq
k_{min}^{-1}$ \cite{Weinberg-Wu}, which turns out to becomparable to
the coherence length of 2SC in accordance with Eq.~(\ref{gapsol}).
(In the case of  homogeneous superconducting phase, $k_{min}=0$, and
$l\rightarrow \infty$.) Considering that the coherence length $\xi$
of a superconductor is proportianl to the inverse of the gap
magnitude, i.e., $\xi \simeq \Delta^{-1}$, therefore, a rather large
ratio of $k_{min}/\Delta$ means a rather small ratio of $l/\xi$.
When $l/\xi <1$, a phase separation state is more favorable.

%%%%%%%%%%%%%%%%%%%%%%%%
%%%%%%%%%%%%%%%%%%%%%%%%
%\hspace{0.1\textwidth}
%\begin{minipage}[t]{0.442\textwidth}
%\end{minipage}
%\end{figure}
%%%%%%%%%%%%%%%%%%%%%%%%%%%%%%%%
%%%%%%%%%%%%%%%%%%%%%%%%%%%%%%%%

The reader has to keep in mind that our result of the Higgs
instability only indicates some kind of inhomogeneous states whose
typical length scale is comparable to the coherence length of 2SC.
Further insight on the structure of the inhomogeneity cannot be
gained without exploring the higher order terms of the nonlinear
realization (\ref{II}). A more direct approach to obtain the
favorite structure of the ground state is to compare the free energy
of various candidate states, which include the mixed phase, the
single-plane wave FF state, striped LO state and multi-plane wave
states. The Coulomb energy and the gradient energy have to be
estimated reliably. We leave this analysis as a future project.

In the system of imbalanced neutral atoms, the Higgs instability
persists and induces spatial non-uniform phase separation state.
This explains why imbalanced cold atom experiments did not observe
LOFF state rather showed strong evidence of phase separation. For
the 2-flavor quark matter being considered,  the electric Coulomb
interaction is not strong enough to compete with the Higgs
instability.

%%%%%%%%%
%%%%%%%%
%%%%%%%%%
%%%%%%%%%

\section{Conclusion}
\label{outlook}

I have introduced several topics of QCD phase structure at high
temperature and high density: the properties of strongly interacting
quark gluon plasma, searching for the critical end point and the
gapless color superconductor.

I give a brief introduction on the discovery of sQGP. It has been
believed for more than 30 years that the QGP created at heavy-ion
collisions should be weakly interacting gas system. However, a very
small shear viscosity over entropy density ratio $\eta/s$ is
required to fit the RHIC data of elliptic flow $v_2$. It is in
contrary to the large value of $\eta/s$ given by perturbative QCD
calculation. This is a strong evidence that the deconfined matter
created at RHIC is strongly interacted. The AdS/CFT duality gives a
lower bound $\eta/s=1/4\pi$. Therefore, it is conjectured that the
sQGP created at RHIC might be the most perfect fluid observed in
nature. However, a perfect fluid should have both vanishing shear
and bulk viscosities. The property of the bulk viscosity of sQGP
need to be studied before one draws the final conclusion.

Recent studies show that the bulk viscosity over entropy density
ratio $\zeta/s$ rises up near phase transitions. The result from QCD
effective models show that $\zeta/s$ behaves differently for
different orders of phase transitions: for 1st-order phase
transition, it rises sharply and show a divergent behavior, for the
2nd-order phase transition, it shows an upward cusp at $T_c$, for
the case of crossover, the cusp becomes smooth. It is discussed that
the sharp rising bulk viscosity will lead to the breakdown of
hydrodynamics and affect the hadronization. Therefore the critical
end point might be located through the observables which are
sensitive to the ratio of bulk viscosity over entropy density.

The status of gapless color superconductor is reported. The
chromomagnetic instability, the Sarma instability and Higgs
instability are clarified. In the gapless color superconducting
phase, both the phase part and magnitude part of the order parameter
will develop instabilities: The phase part develops into the
chromomagnetic instability, which induces the plane-wave state; The
magnitude part develops the Sarma instability and Higgs instability,
the Sarma instability can be competed with charge neutrality
condition, while the Higgs instability cannot be cured by Coulomb
interaction, and induces the inhomogeneous state.

\section{Acknowledgements}
I thank the collaboration with J.W.Chen, B.C.Li, H.Mao, D.L.Yang for
the sQGP part, and I.Giannakis, D.Hou, H.C.Ren and I. Shovkovy for
the gapless color superconductor part. The work is supported by CAS
program "Outstanding young scientists abroad brought-in", CAS key
project KJCX3-SYW-N2, NSFC10735040, NSFC10875134, and the K.C.Wong
Education Foundation, Hong Kong.

\end{document}